\documentclass[floats, prd, eqnum, showpacs, nofootinbib, 
twocolumn, 
eqsecnum]{revtex4-1}

\usepackage{color,graphicx}
\usepackage{amsfonts}
\usepackage{amssymb}
\usepackage{comment}
\begin{document}

\title{Nonlogarithmic divergence of a deflection angle by a marginally unstable photon sphere of the Damour-Solodukhin wormhole in a strong deflection limit}
\author{Naoki Tsukamoto${}^{1}$}\email{tsukamoto@rikkyo.ac.jp}
\affiliation{
${}^{1}$Department of General Science and Education, National Institute of Technology, Hachinohe College, Aomori 039-1192, Japan \\
}

\begin{abstract}
Static, spherically symmetric black holes and compact objects without an event horizon have unstable (stable) circular orbits of a light called photon (antiphoton) sphere.
A Damour-Solodukhin wormhole has been suggested as a simple black hole mimicker and the difference of its metric tensors from a black hole is described by a dimensionless parameter $\lambda$.
The wormhole with two flat regions has two photon spheres and an antiphoton sphere for $\lambda<\sqrt{2}/2$ 
and a photon sphere for $\lambda \geq \sqrt{2}/2$.
When the parameter $\lambda$ is $\sqrt{2}/2$, the photon sphere is marginally unstable because of degeneration of the photon spheres and antiphoton sphere.
We investigate gravitational lensing by the wormhole in weak and strong gravitational fields.
We find that the deflection angle of a light ray reflected by the marginally unstable photon sphere diverges nonlogarithmically in a strong deflection limit for $\lambda=\sqrt{2}/2$,
while the deflection angle reflected by the photon sphere diverges logarithmically for $\lambda \neq \sqrt{2}/2$.
We extend a strong deflection limit analysis for the nonlogarithmic divergence case.
We expect that our method can be applied for gravitational lenses by marginally unstable photon spheres of various compact objects.
\end{abstract}

\maketitle

\section{Introduction}
Recently, LIGO and VIRGO Collaborations have reported the direct detection of gravitational waves from black holes~\cite{Abbott:2016blz,LIGOScientific:2018mvr} and
Event Horizon Telescope Collaboration has reported the ring image of supermassive black hole candidates at the center of a giant elliptical galaxy M87~\cite{Akiyama:2019cqa}.
The black holes and the other compact objects with a strong gravitational field described by general relativity will be more important to understand our universe. 

It is well known that static, spherically symmetric compact objects have unstable (stable) circular photon orbit called photon (antiphoton) sphere~\cite{Claudel:2000yi,Perlick_2004_Living_Rev}.
The upper bound of the radius of the (anti)photon sphere of the static, spherically symmetric black hole under the weak energy condition is given by $r=3M$, where $M$ is the mass 
of the black hole~\cite{Hod:2017xkz}.  
The (anti)photon sphere has important roles in several phenomena in a strong gravitational field:
Light rays emitted by a source and reflected by the photon sphere make infinite number of dim images~\cite{Hagihara_1931,Darwin_1959,Atkinson_1965,Luminet_1979,Ohanian_1987,Nemiroff_1993,Frittelli_Kling_Newman_2000,Virbhadra_Ellis_2000,Bozza_Capozziello_Iovane_Scarpetta_2001,Bozza:2002zj,Perlick_2004_Phys_Rev_D,Bozza_2010}, 
which are named relativistic images in Ref.~\cite{Virbhadra_Ellis_2000}, on the both sides of the photon sphere and we can survey the compact object with the photon sphere by the images even if the compact objects themselves do not emit light rays.
The photon sphere can be observed during a collapsing star to be a black hole~\cite{Ames_1968,Synge:1966okc,Yoshino:2019qsh}.
The photon sphere has strong influence on the high-frequency behavior of the photon absorption cross section~\cite{Sanchez:1977si,Decanini:2010fz} 
and the high-frequency spectrum of quasinormal modes of compact objects~\cite{Press:1971wr,Goebel_1972,Raffaelli:2014ola}.
An observer moving on the photon sphere feels no centrifugal force and no gyroscopic precession~\cite{Abramowicz_Prasanna_1990,Abramowicz:1990cb,Allen:1990ci,Hasse_Perlick_2002}
and it is the fastest way to circle a static, spherically symmetric black hole for massless particles~\cite{Hod:2012nk}.
The photon sphere is correspond to Bondi's sonic horizon of a radial fluid~\cite{Mach:2013gia,Chaverra:2015bya,Cvetic:2016bxi,Koga:2016jjq,Koga:2018ybs,Koga:2019teu}.

Stability of light rings, i.e., circular photon orbits~\cite{Koga:2019uqd}, of compact objects are an important property of the spacetime.
Instability of the compact objects with the stable light rings has been concerned since they cause the slow decay of linear waves~\cite{Keir:2014oka,Cardoso:2014sna,Cunha:2017qtt}.
The numbers of light rings of stationary, axisymmetic compact objects without an event horizon under energy conditions are two at least, they are even number in general, and the inner light ring is stable~\cite{Cunha:2017eoe}.
Hod has shown that spherically symmetric compact objects without an event horizon have odd number light rings because of degeneration~\cite{Hod:2017zpi}. 
We note that we cannot apply the theorem of the number of the light rings for wormholes since a trivial topology has been assumed in Refs.~\cite{Cunha:2017eoe,Hod:2017zpi}.

General relativity permits the wormholes which can have nontrivial topological structures~\cite{Visser_1995,Morris_Thorne_1988}.
The wormholes have a throat which connects two regions of one universe or two universes.
Gravitational lensing~\cite{Schneider_Ehlers_Falco_1992,Schneider_Kochanek_Wambsganss_2006,Perlick_2004_Living_Rev} is used to find dark gravitating objects like wormholes.
However, we cannot distinguish the wormhole with a positive mass from other massive objects such as a black hole under a weak-field approximation. 
Therefore, we must rely on the observation near the throat or the photon sphere in a strong gravitational field  such as 
gravitational lensing of light rays scatter by the throat or the photon sphere~\cite{Chetouani_Clement_1984,Perlick_2004_Phys_Rev_D,Nandi:2006ds,Muller:2008zza,Tsukamoto_Harada_Yajima_2012,Perlick:2014zwa,Tsukamoto:2016qro,Tsukamoto:2016zdu,Nandi:2016uzg,Tsukamoto:2017edq,Shaikh:2018oul,Shaikh:2019jfr,Shaikh:2019itn},
visualizations~\cite{Muller_2004,James:2015ima}, 
shadows in an accretion gas~\cite{Ohgami:2015nra,Ohgami:2016iqm,Kuniyasu:2018cgv}, 
wave optics~\cite{Nambu:2019sqn},
and gravitational waves~\cite{Cardoso:2016rao}
to distinguish wormhole from the other compact objects.

A Damour-Solodukhin wormhole~\cite{Damour:2007ap} has been
suggested as a black hole mimicker.
Its metric was created by making a slight modification to the Schwarzschild metric and 
it can be the simplest metric among black hole mimickers.
The difference of the metric from the Schwarzschild spacetime is described by a positive dimensionless parameter~$\lambda$.
Lemos and Zaslavskii have pointed out that we can distinguish the wormhole from the black hole by a tidal force acting on a body near the throat even if the difference of the metric tensors of the wormhole from the black hole is small, i.e., $\lambda \ll 1$~\cite{Lemos:2008cv}. 
Emissions from its accretion disk~\cite{Karimov:2019qco},
quasinormal modes and gravitational waves~\cite{Bueno:2017hyj,Volkel:2018hwb},
images of its accretion disks~\cite{Paul:2019trt}, 
shadow~\cite{Amir:2018pcu},
gravitational lensing~\cite{Nandi:2018mzm,Ovgun:2018fnk,Bhattacharya:2018leh,Ovgun:2018swe,Ovgun:2018oxk}, 
and particle collision~\cite{Tsukamoto:2019ihj} 
in the Damour-Solodukhin wormhole spacetime have been investigated.

Gravitational lensing in a strong deflection limit, which is a semianalytic formalism for gravitational lensing of a light ray reflected by the photon sphere, 
has been investigated by Bozza~\cite{Bozza:2002zj}.
The deflection angle $\alpha_{\mathrm{def}}$ of the light ray in the strong deflection limit has the following form 
\begin{eqnarray}\label{eq:deflection_angle_limit_0} 
\alpha_{\mathrm{def}}(b)&=& -\bar{a} \log \left( \frac{b}{b_{\mathrm{m}}}-1 \right) + \bar{b} \nonumber\\
&&+O\left( \left( \frac{b}{b_{\mathrm{m}}}-1 \right) \log \left( \frac{b}{b_{\mathrm{m}}}-1 \right)\right)
\end{eqnarray}
or 
\begin{eqnarray}\label{eq:deflection_angle_limit_0th} 
\alpha_{\mathrm{def}}(\theta)&=& -\bar{a} \log \left( \frac{\theta}{\theta_{\infty}}-1 \right) + \bar{b} \nonumber\\
&&+O\left( \left( \frac{\theta}{\theta_{\infty}}-1 \right) \log \left( \frac{\theta}{\theta_{\infty}}-1 \right)\right),
\end{eqnarray}
where $b$ and $b_{\mathrm{m}}$ are the impact parameter and the critical impact parameter of the light ray, respectively,
$\theta$ is an image angle, 
$\theta_{\infty}\equiv b_{\mathrm{m}}/D_{OL}$ is the image angle of the photon sphere, where $D_{OL}$ is a distance between an observer and a lens object,
and $\bar{a}$ is a positive parameter and $\bar{b}$ is a parameter.
We can assume that the impact parameter $b$ is non-negative without loss of generality when we treat one light ray in a spherically symmetric spacetime.
The analysis in the strong deflection limit has been improved~\cite{Eiroa:2002mk,Bozza:2005tg,Bozza:2006nm,Bozza:2007gt,Tsukamoto:2016qro,Tsukamoto:2016jzh,Ishihara:2016sfv,Shaikh:2019itn,Shaikh:2019jfr}
and its relations to high-energy absorption cross section~\cite{Wei:2011zw} and to quasinormal modes~\cite{Stefanov:2010xz} have been investigated.
Recently, 
Shaikh \textit{et al.} have considered the deflection angle of light rays scattered by a photon sphere at the throat of a wormhole~\cite{Shaikh:2018oul,Shaikh:2019jfr}. 

The deflection angle of a light ray scattered by a photon sphere of the Damour-Solodukhin wormhole 
in the strong deflection limit was obtained by Nandi~\textit{et al.}~\cite{Nandi:2018mzm}, Ovgun~\cite{Ovgun:2018fnk}, and Bhattacharya and Karimov~\cite{Bhattacharya:2018leh}
when the wormhole metric is similar to the black hole, i.e., for $\lambda < \sqrt{2}/2$. 
In Ref.~\cite{Bhattacharya:2018leh}, Bhattacharya and Karimov have pointed out that $\bar{b}$ by Ovgun~\cite{Ovgun:2018fnk} is in error. 

In this paper, firstly we reexamine the deflection angle of the light ray scattered by the photon spheres of the Damour-Solodukhin wormhole for $\lambda < \sqrt{2}/2$.
We recover Bhattacharya and Karimov's result~\cite{Bhattacharya:2018leh} and we give a small modification for calculation in \cite{Nandi:2018mzm}.
Secondly we extend the analysis for $\lambda \geq \sqrt{2}/2$.
The deflection angles diverge logarithmically in the strong deflection limit if the dimensionless parameter $\lambda \neq \sqrt{2}/2$.
Interestingly, we have found that the deflection angle of a light ray scattered by a marginally unstable photon sphere at the throat diverges nonlogarithmically for $\lambda=\sqrt{2}/2$.
We construct the strong deflection limit analysis for the deflection angle with a nonlogarithmic divergence in the Damour-Solodukhin wormhole spacetime.

This paper is organized as follows. 
In Sec.~II, we review the Damour-Solodukhin wormhole spacetime. 
In Secs.~III and~IV, we investigate the deflection angle and observables, respectively, in the strong deflection limit.
We review the gravitational lensing under a weak-field approximation in Sec.~V
and we summarize our result in Sec.~VI.
In Appendixes~A and B, the Arnowitt-Deser-Misner (ADM) masses and the violation of energy conditions of the Damour-Solodukhin wormhole are shown, respectively.
In this paper we use the units in which a light speed and Newton's constant are unity.

\section{Damour-Solodukhin wormhole spacetime}
In this section, we review the trajectory of a light ray and its deflection angle $\alpha_{\mathrm{def}}$ in a Damour-Solodukhin wormhole spacetime~\cite{Damour:2007ap} with a line element given by, in coordinates $(\tilde{t}, r, \vartheta, \varphi)$,
\begin{eqnarray}\label{eq:line_element} 
ds^2
&=&-\left( 1-\frac{2\tilde{M}}{r}+\lambda^2 \right)d\tilde{t}^2 +\frac{dr^2}{1-\frac{2\tilde{M}}{r}} \nonumber\\
&&+r^2 \left( d \vartheta^2 +\sin^2\vartheta d \varphi^2 \right),
\end{eqnarray}
where $\lambda$ and $\tilde{M}$ are positive parameters and the radial coordinate $r$ is defined in a range $2\tilde{M} \leq r <\infty$.
A throat exists at $r=r_{\mathrm{th}}\equiv 2\tilde{M}$. 
The wormhole spacetime can take different values of the parameters $\tilde{M}$ and $\lambda$ in two asymptotically flat regions. 
For simplicity, we assume the equal parameters in the both regions.   
The line element is the same the one in the Schwarzschild spacetime in a limit~$\lambda \rightarrow 0$ and
it is the same as the static case of a Kerr-like wormhole~\cite{Bueno:2017hyj}. 
Bozza has considered the deflection angle of a light ray in the strong deflection limit in a general asymptotically flat, static, and spherically symmetric spacetime
with its metric tensor behaving
\begin{eqnarray}\label{eq:metric_assumption_tt} 
&&\lim_{r\rightarrow \infty}g_{tt} \rightarrow -\left(1+\frac{2M}{r} \right), \\ \label{eq:metric_assumption_rr}
&&\lim_{r\rightarrow \infty}g_{rr} \rightarrow 1-\frac{2M}{r}, \\
&&\lim_{r\rightarrow \infty}g_{\vartheta\vartheta}=\lim_{r\rightarrow \infty}\frac{g_{\varphi \varphi}}{\sin^2 \vartheta} \rightarrow r^2,
\end{eqnarray}
where $t$ is a time coordinate and $M$ is a positive parameter~\cite{Bozza:2002zj}.
The Damour-Solodukhin wormhole spacetime is an asymptotically flat, static, and spherically symmetric spacetime 
but the $(\tilde{t},\tilde{t})$-component of the metric tensor $g_{\tilde{t}\tilde{t}}$~(\ref{eq:line_element}) does not satisfy the assumption~(\ref{eq:metric_assumption_tt}). 
We notice that $(\tilde{t},\tilde{t})$-component of the metric tensor $g_{\tilde{t}\tilde{t}}$ shown in Eq.~(\ref{eq:line_element}) behaving asymptotically
\begin{eqnarray}
\lim_{r\rightarrow \infty}g_{\tilde{t}\tilde{t}} \rightarrow -(1+\lambda^2)
\end{eqnarray} 
is not suitable for a variable $z$ defined in Sec.~III.
Thus, we introduce a new time coordinate~$t$ and a positive parameter $M$ which are defined by
\begin{eqnarray}
t\equiv \frac{\tilde{t}}{\sqrt{1+\lambda^2}}
\end{eqnarray}
and 
\begin{eqnarray}
M \equiv \frac{\tilde{M}}{1+\lambda^2}, 
\end{eqnarray} 
respectively, to satisfy the assumption~(\ref{eq:metric_assumption_tt}).
By using $t$ and $M$, the line element~(\ref{eq:line_element}) is rewritten in
\begin{eqnarray}\label{eq:line_element2} 
ds^2=-A(r)dt^2 +B(r)dr^2+C(r) \left( d \vartheta^2 +\sin^2\vartheta d \varphi^2 \right), \nonumber\\
\end{eqnarray}
where $A(r)$, $B(r)$, and $C(r)$ are given by 
\begin{eqnarray}\label{eq:Metric} 
&&A(r) \equiv 1-\frac{2M}{r}, \\ \label{eq:Metric_B} 
&&B(r) \equiv \left[1-\frac{2M(1+\lambda^2)}{r} \right]^{-1}, 
\end{eqnarray}
and 
\begin{eqnarray}
C(r) \equiv r^2,
\end{eqnarray}
respectively.~\footnote{The metric tensor~$(\ref{eq:line_element2})$ does not satisfy the condition~(\ref{eq:metric_assumption_rr}) but it does not give us troubles to define the variable $z$.} 
Notice that $r_{\mathrm{th}}$ is rewritten as $r_{\mathrm{th}}=2M(1+\lambda^2)$.
Since the spacetime is a static, spherically symmetric spacetime, 
there are time-translational and axial Killing vectors 
$t^\mu \partial_\mu=  \partial_t$ and $\varphi^\mu \partial_\mu = \partial_\varphi$, respectively. 

From $k^\mu k_\mu=0$, where $k^\mu \equiv \dot{x}^\mu$ is the wave number of a light ray and where the dot denotes the differentiation with respect to an affine parameter,
the trajectory of the light ray is obtained as 
\begin{eqnarray}\label{eq:trajectory} 
-A(r)\dot{t}^2+B(r)\dot{r}^2+C(r)\dot{\varphi}^2=0.
\end{eqnarray}
Here we have set $\vartheta=\pi/2$ without loss of generality.
Equation~(\ref{eq:trajectory}) can be expressed as
\begin{eqnarray}
\dot{r}^2+\tilde{V}(r)=0,
\end{eqnarray}
where the effective potential $\tilde{V}(r)$ is defined by
\begin{eqnarray}
\tilde{V}(r)\equiv -\frac{E^2}{B(r)}\left( \frac{1}{A(r)}-\frac{b^2}{C(r)} \right),
\end{eqnarray}
where $b\equiv L/E$ is the impact parameter of the light ray and the conserved energy $E\equiv -g_{\mu \nu}t^{\mu}k^{\nu}$ and the conserved angular momentum $L\equiv g_{\mu \nu}\varphi^{\mu}k^{\nu}$ of the light ray
are constant along the trajectory. Since $\tilde{V}(r) \rightarrow E^2>0$ in spatial infinity $r \rightarrow \infty$, the light ray exists there.

By introducing a proper radial distance $l$ from the throat given by
\begin{eqnarray}
l&\equiv& \int^{r}_{r_{\mathrm{th}}}\sqrt{B(r)}dr \nonumber\\
&=&\sqrt{r^2 - 2M \left(1+\lambda^2 \right) r } \nonumber\\
&&+ M \left(1+ \lambda ^2 \right)  \log \frac{r-M(1+\lambda^2)+\sqrt{r^2-2M(1+\lambda^2)r}}{M(1+\lambda^2)}, \nonumber\\
\end{eqnarray}
the line element, the equation of trajectory of the light ray, and the effective potential are rewritten in
\begin{eqnarray}\label{eq:line_element3} 
ds^2
=-\left( 1-\frac{2M}{r(l)} \right)dt^2 +dl^2 +r^2(l) \left( d \vartheta^2 +\sin^2\vartheta d \varphi^2 \right), \nonumber\\
\end{eqnarray}
\begin{eqnarray}
\dot{l}^2+\tilde{v}(l)=0,
\end{eqnarray}
and 
\begin{eqnarray}\label{eq:tilde_v}
\tilde{v}(l)\equiv -E^2 \left( \frac{1}{A(l)}-\frac{b^2}{C(l)} \right),
\end{eqnarray}
respectively.
Note that the proper radial distance $l$ is defined in a range $-\infty <l < \infty$ and the throat is at $l=0$.

From $\tilde{v}=\frac{d\tilde{v}}{dl}=0$,
we get the circular light orbits with $b=2M(1+\lambda^2)^{\frac{3}{2}}/\lambda$ and $b=3\sqrt{3}M$ at $r=r_{\mathrm{th}}=2M(1+\lambda^2)$ and $r=3M$, respectively.
From straightforward calculations, 
we obtain $\left. \frac{d^2\tilde{v}}{dl^2} \right|_{r=r_{\mathrm{th}}}>0$ for $\lambda<\sqrt{2}/2$ 
and $\left. \frac{d^2\tilde{v}}{dl^2} \right|_{r=r_{\mathrm{th}}}\leq 0$ for $\lambda \geq  \sqrt{2}/2$. 
Therefore, the throat is a photon (antiphoton) sphere for $\lambda>\sqrt{2}/2$ ($\lambda<\sqrt{2}/2$) and the photon sphere is marginally unstable for $\lambda=\sqrt{2}/2$.
Figure~1 shows the dimensionless effective potential $v\equiv \tilde{v}/E^2$ for the (marginally unstable) photon sphere and antiphoton sphere.
\begin{figure}[htbp]
\begin{center}
\includegraphics[width=87mm]{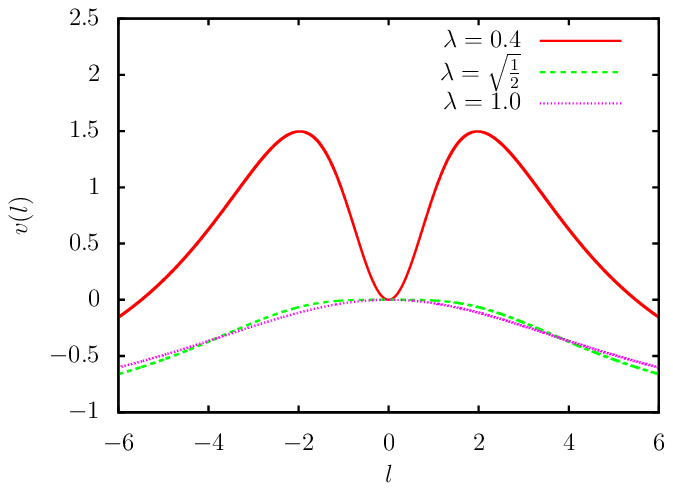}
\end{center}
\caption{A dimensionless effective potential $v(l)$ as a function of a proper distance $l$ from the throat. It shows a (anti)photon sphere at the throat.
The solid~(red), broken~(green), and dotted~(magenta) curves denote the effective potential $v(l)$ when $\lambda=0.4$, $\sqrt{2}/2$, and $1$, respectively.
We have set $M=1$ and $b=2M(1+\lambda^2)^{\frac{3}{2}}/\lambda$.}
\end{figure}
The circular orbit of a light ray with $b=3\sqrt{3}$ at $r=3M$ is unstable for $\lambda \leq \sqrt{2}/2$, i.e., it is a photon sphere.
Notice that the wormhole has two asymptotically flat regions and it has two photon spheres at $r=3M$ for $\lambda < \sqrt{2}/2$. 
The photon spheres at $r=3M$ and the antiphoton sphere at the throat $r=r_{\mathrm{th}}=2M(1+\lambda^2)$ 
degenerate to be a marginally unstable photon sphere at the throat just for $\lambda=\sqrt{2}/2$.
The wormhole has only one photon sphere at the throat for $\lambda > \sqrt{2}/2$. 

As shown Fig.~2,
we can classify the light ray which coming from a spatial infinity into a falling case with $b<b_{\mathrm{m}}$, a critical case with $b=b_{\mathrm{m}}$, and a scattered case with $b<b_{\mathrm{m}}$. 
Here the critical impact parameter $b_{\mathrm{m}}$ is defined by
\begin{eqnarray}\label{eq:b_m} 
b_{\mathrm{m}}\equiv \lim_{r_0 \rightarrow r_{\mathrm{m}}} b(r_0)= \sqrt{\frac{C_{\mathrm{m}}}{A_{\mathrm{m}}}},
\end{eqnarray}
where $r_0$ is the closest distance of the light ray 
and hereafter subscript $m$ denotes quantities of the photon sphere at $r=r_{\mathrm{m}}$.
\begin{figure}[htbp]
\begin{center}
\includegraphics[width=87mm]{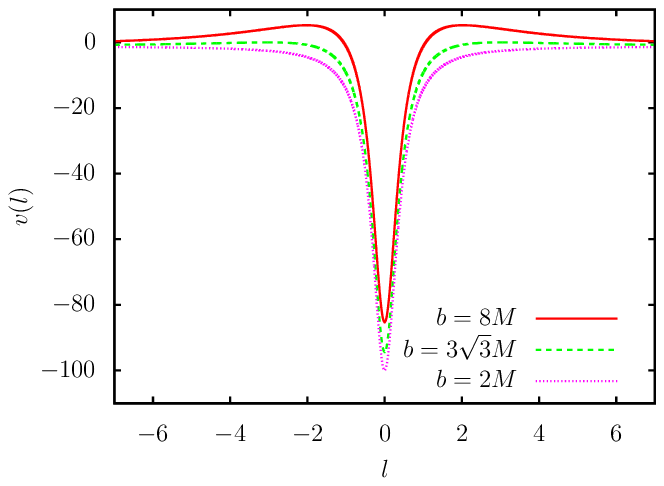}
\includegraphics[width=87mm]{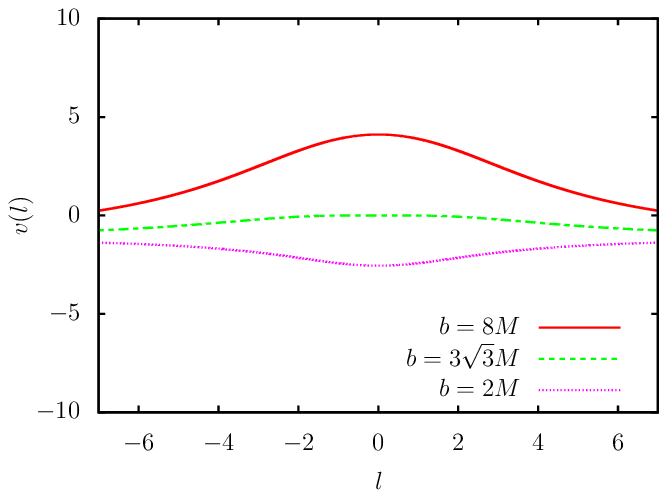}
\end{center}
\caption{A dimensionless effective potential $v(l)$ as a function of a proper distance $l$ from the throat.
The solid~(red), broken~(green), and dotted~(magenta) curves denote the effective potential $v(l)$ in scattered $(b=8M)$, critical $(b=3\sqrt{3}M)$, and falling $(b=2M)$ cases, respectively.
The upper and lower panels show cases with $\lambda=0.1$ and $\lambda=\sqrt{2}/2$, respectively. All the cases, we have set $M=1$.
The effective potential $v(l)$ for $\lambda > \sqrt{2}/2$ has a quietly similar shape to the one for $\lambda = \sqrt{2}/2$.}
\end{figure}

We concentrate on the scattered case. In this case, a light ray comes from a spatial infinity, 
it is deflected by the wormhole at a reflection point $r=r_0$,  
and it goes back to the same spatial infinity. 
The reflection point is obtained as the largest positive solution of the equation $v(l)=0$ or $V(r)=0$.
At the reflection point $r=r_0$, the equation of the trajectory~(\ref{eq:trajectory}) gives
\begin{eqnarray}\label{eq:quantity_0} 
A_0\dot{t}^2_0=C_0\dot{\varphi}^2_0,
\end{eqnarray}
where the subscript~$0$ denotes the quantity at $r=r_0$.
The impact parameter $b=b(r_0)$ is expressed by
\begin{eqnarray}\label{eq:b_0} 
b(r_0)=\frac{L}{E}=\frac{C_0 \dot{\varphi}_0}{A_0 \dot{t}_0}=\sqrt{\frac{C_0}{A_0}}.
\end{eqnarray}
Here we have used Eq.~(\ref{eq:quantity_0}). 

From Eq.~(\ref{eq:trajectory}), the deflection angle $\alpha_{\mathrm{def}}(r_0)$ of the light ray as a function of the reflection point $r_0$ is obtained as
\begin{eqnarray}
\alpha_{\mathrm{def}}(r_0)\equiv I(r_0)-\pi,
\end{eqnarray}
where $I(r_0)$ is defined by
\begin{eqnarray}
I(r_0)&\equiv& 2\int^{\infty}_{r_0} \frac{\sqrt{B(r)}dr}{\sqrt{C(r)}\sqrt{\frac{C(r)A_0}{A(r)C_0}-1}} \nonumber\\
&=& 2\int^{\infty}_{r_0} \frac{b dr}{C(r)\sqrt{-V(r)}} \nonumber\\
&=& 2\int^{\infty}_{l_0} \frac{b dl}{C(l)\sqrt{-v(l)}},
\end{eqnarray}
where $V(r)\equiv \tilde{V}(r)/E^2$ is a dimensionless effective potential in the radial coordinate $r$ 
and where $l_0\equiv \left. l(r) \right|_{r=r_0}$ is the position of the reflection point in the radial coordinate $l$.

\section{Deflection angle in a strong deflection limit}
In this section, we investigate the deflection angle in the strong deflection limit $r_0 \rightarrow r_{\mathrm{m}}$ or $b \rightarrow b_{\mathrm{m}}$.
We treat it in the cases for $\lambda<\sqrt{2}/2$, $\lambda>\sqrt{2}/2$, and $\lambda=\sqrt{2}/2$ in this order. 
In the strong deflection limit $b \rightarrow b_{\mathrm{m}}$,
the deflection angle of the light ray in a strong deflection is expressed by a following form~\cite{Bozza:2002zj}:~\footnote{The subleading term is $O(b-b_{\mathrm{m}})$ in Ref.~\cite{Bozza:2002zj} 
but we should read it as $O\left(\left( \frac{b}{b_{\mathrm{m}}}-1 \right) \log \left( \frac{b}{b_{\mathrm{m}}}-1 \right)\right)$. See Refs.~\cite{Tsukamoto:2016jzh,Tsukamoto:2016qro,Iyer:2006cn}.}             
\begin{eqnarray}\label{eq:deflection_angle_limit} 
\alpha_{\mathrm{def}}(b)&=& -\bar{a} \log \left( \frac{b}{b_{\mathrm{m}}}-1 \right) + \bar{b} \nonumber\\
&&+O\left( \left( \frac{b}{b_{\mathrm{m}}}-1 \right) \log \left( \frac{b}{b_{\mathrm{m}}}-1 \right)\right). 
\end{eqnarray}
We introduce a variable $z$~\cite{Bozza:2002zj} defined by
\begin{eqnarray}
z\equiv \frac{A(r)-A_0}{1-A_0}=1-\frac{r_0}{r}.
\end{eqnarray}

\subsection{$\lambda<\sqrt{2}/2$}
In the case for $\lambda<\sqrt{2}/2$, 
the photon sphere is at $r=r_{\mathrm{m}}=3M$
and the critical impact parameter $b_{\mathrm{m}}$ is given by $b_{\mathrm{m}}=3\sqrt{3}M$.
By using $z$, we rewrite $I(r_0)$ as
\begin{eqnarray}
I(r_0)= \int^{1}_{0} R(z,r_0)f(z,r_0)dz,
\end{eqnarray}
where $R(z,r_0)$ is given by
\begin{eqnarray}
R(z,r_0)
&\equiv& \frac{2\sqrt{AB}}{A'C}(1-A_0)\sqrt{C_0} \nonumber\\
&=&2\sqrt{\frac{r_0-2M(1-z)}{r_0-2M(1+\lambda^2)(1-z)}},   
\end{eqnarray}
where $'$ is the differentiation with respect to $r$ and $f(z,r_0)$ is given by
\begin{eqnarray}
f(z,r_0)
&\equiv& \frac{1}{\sqrt{A_0-\left[ (1-A_0)z+A_0 \right]\frac{C_0}{C}}} \nonumber\\
&=&\frac{1}{\sqrt{\alpha_0 z+\beta_0 z^2 -\frac{2M}{r_0}z^3}}, \nonumber\\
\end{eqnarray}
and $\alpha_0$ and $\beta_0$ are defined by
\begin{eqnarray}
&&\alpha_0=\alpha (r_0) \equiv  2-\frac{6M}{r_0},\\
&&\beta_0=\beta (r_0) \equiv -1+\frac{6M}{r_0}.
\end{eqnarray}
$R(z,r_0)$ is regular but $f(z,r_0)$ diverges in a limit $z\rightarrow 0$. 
We define $f_0(z,r_0)$ as
\begin{eqnarray}
f_0(z,r_0) \equiv \frac{1}{\sqrt{\alpha_0 z +\beta_0 z^2}}.
\end{eqnarray}
Since $\alpha_{\mathrm{m}}=0$ and $\beta_{\mathrm{m}}=1$, the integral of $f_0(z,r_0)$ diverges in the strong deflection limit $r_0 \rightarrow r_{\mathrm{m}}$.
By using $f_0(z,r_0)$, we separate $I(r_0)$ into a divergent part $I_{\mathrm{D}}(r_0)$ and a regular part $I_{\mathrm{R}}(r_0)$:
\begin{eqnarray}
I(r_0)=I_{\mathrm{D}}(r_0)+I_{\mathrm{R}}(r_0),
\end{eqnarray}
where 
\begin{eqnarray}
&&I_{\mathrm{D}}(r_0)=\int^{1}_{0} R(0,r_{\mathrm{m}})f_0(z,r_0) dz, \\
&&I_{\mathrm{R}}(r_0)=\int^{1}_{0} g(z,r_0) dz, 
\end{eqnarray}
where $g(z,r_0)$ is defined by
\begin{eqnarray}
g(z,r_0)\equiv R(z,r_0)f(z,r_0)-R(0,r_{\mathrm{m}})f_0(z,r_0).
\end{eqnarray}

The divergent part $I_{\mathrm{D}}(r_0)$ can be integrated and it becomes
\begin{eqnarray}\label{eq:I_D_0} 
&&I_{\mathrm{D}}(r_0)=  \frac{2R(0,r_{\mathrm{m}})}{\sqrt{\beta_0}} \log \frac{\sqrt{\beta_0}+\sqrt{\alpha_0+\beta_0}}{\sqrt{\alpha_0}}.
\end{eqnarray}
We expand $\alpha_0$ and $\beta_0$ in powers of $(r_0-r_{\mathrm{m}})$:
\begin{eqnarray}
&&\alpha_0=\frac{2}{3M}(r_0-r_{\mathrm{m}}) +O\left((r_0-r_{\mathrm{m}})^2\right),\\
&&\beta_0=1+ O \left( r_0-r_{\mathrm{m}} \right).
\end{eqnarray}
By substituting them into Eq.~(\ref{eq:I_D_0}), we obtain
\begin{eqnarray}
I_{\mathrm{D}}(r_0)&=&-\frac{2}{\sqrt{1-2\lambda^2}} \log \left( \frac{r_0}{r_{\mathrm{m}}}-1 \right) +\frac{2\log 2}{\sqrt{1-2\lambda^2}} \nonumber\\
&&+O\left( \left( \frac{r_0}{r_{\mathrm{m}}}-1 \right) \log \left( \frac{r_0}{r_{\mathrm{m}}}-1 \right)\right). 
\end{eqnarray}
By using the impact parameter $b$ expanded in powers of~$(r_0-r_{\mathrm{m}})$ obtained as
\begin{eqnarray}\label{eq:b_r} 
b=b_{\mathrm{m}}+\frac{\sqrt{3}}{2M}(r_0-r_{\mathrm{m}})^2 +O\left( (r_0-r_{\mathrm{m}})^3 \right),
\end{eqnarray}
we can rewrite $I_{\mathrm{D}}=I_{\mathrm{D}}(b)$ as
\begin{eqnarray}
I_{\mathrm{D}}(b)&=&-\frac{1}{\sqrt{1-2\lambda^2}} \log \left( \frac{b}{b_{\mathrm{m}}}-1 \right) +\frac{\log 6}{\sqrt{1-2\lambda^2}} \nonumber\\
&&+O\left( \left( \frac{b}{b_{\mathrm{m}}}-1 \right) \log \left( \frac{b}{b_{\mathrm{m}}}-1 \right)\right). 
\end{eqnarray}

The regular part $I_{\mathrm{R}}(r_0)$ can be expanded in powers of $(r_0-r_{\mathrm{m}})$ and it is expressed by
\begin{eqnarray}
I_{\mathrm{R}}(r_0)=\sum^{\infty}_{j=0} \frac{1}{j!} (r_0-r_{\mathrm{m}})^j \int^1_0 \left. \frac{\partial^j g}{\partial r^j_0} \right|_{r_0=r_{\mathrm{m}}} dz.
\end{eqnarray}
We are interested in $j=0$ term:
\begin{eqnarray}
I_{\mathrm{R}}(r_0)
&=&\int^1_0 g(z, r_{\mathrm{m}})dz \nonumber\\
&&+O\left( \left( \frac{r_0}{r_{\mathrm{m}}}-1 \right) \log \left( \frac{r_0}{r_{\mathrm{m}}}-1 \right)\right), 
\end{eqnarray}
where $g(z, r_{\mathrm{m}})$ is given by
\begin{eqnarray}
g(z, r_{\mathrm{m}})
&=&\frac{2\sqrt{3+6z}}{z\sqrt{1-2\lambda^2+2(1+\lambda^2)z} \sqrt{3-2z}} \nonumber\\
&&-\frac{2}{z\sqrt{1-2\lambda^2}}.
\end{eqnarray}

Thus, we have obtained the deflection angle $\alpha_{\mathrm{def}}(b)$ of the light ray in the strong deflection limit $b\rightarrow b_{\mathrm{m}}$ in the form of Eq.~(\ref{eq:deflection_angle_limit}) 
with 
\begin{eqnarray}
\bar{a}&=&\frac{1}{\sqrt{1-2\lambda^2}},\\ \label{eq:b_our} 
\bar{b}&=&\frac{\log 6}{\sqrt{1-2\lambda^2}} +I_{\mathrm{R}} -\pi.
\end{eqnarray}
This is the same as the deflection angle obtained by Bhattacharya and Karimov~\cite{Bhattacharya:2018leh}.
We plot $\bar{a}$ and $\bar{b}$ for $\lambda \neq \sqrt{2}/2$ in Fig.~3.
\begin{figure}[htbp]
\begin{center}
\includegraphics[width=87mm]{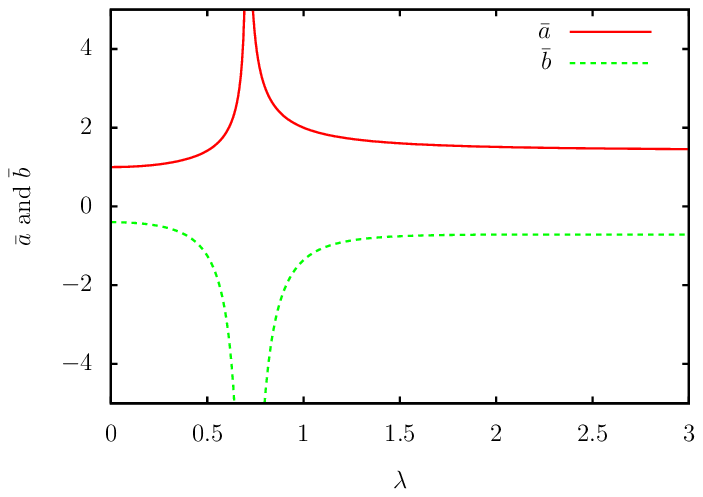}
\end{center}
\caption{$\bar{a}$ and $\bar{b}$. 
The solid~(red) and dashed~(green) curves denote $\bar{a}$ and $\bar{b}$, respectively.
In a limit $\lambda \rightarrow \sqrt{2}/2$, we obtain $\bar{a} \rightarrow \infty$ and $\bar{b} \rightarrow -\infty$.}
\end{figure}

We comment on the details of earlier work on it.
First, the deflection angle in the strong deflection limit was calculated by Nandi~\textit{et al.}~\cite{Nandi:2018mzm}.
They considered the metric tensor~(\ref{eq:line_element}) in the coordinates $(\tilde{t}, r, \vartheta, \varphi)$
and they defined the variable~\footnote{Note that we use $(-,+,+,+)$, while Nandi~\textit{et al.} have used $(+,-,-,-)$~\cite{Nandi:2018mzm}.}
\begin{eqnarray}
z_{[74]}
\equiv \frac{g_{\tilde{t} \tilde{t}}(r)-g_{\tilde{t} \tilde{t}}(r_0)}{1-g_{\tilde{t} \tilde{t}}(r_0)} 
=\frac{2\tilde{M}}{2\tilde{M}-\lambda^2 r_0} \left( 1-\frac{r_0}{r} \right).
\end{eqnarray}
Note that 
the integration range of Eq.~(2.10) in Ref.~\cite{Nandi:2018mzm},  $0 \leq z_{[74]} \leq 1$ should be modified to be 
\begin{eqnarray}
0 \leq z_{[74]} \leq \frac{2\tilde{M}}{2\tilde{M}-\lambda^2 r_0}.  
\end{eqnarray}
The error affects to $\bar{b}$ and the difference is small when $\lambda$ is small as shown Table I. 
Their results in Ref.~\cite{Nandi:2018mzm} will be valid since they discussed in the case of $\lambda \ll 1$.
\begin{table}[htbp]
\label{table:I}
\caption{Parameters $\bar{a}$ and $\bar{b}$ as a function with respect to $\lambda$. We have obtained the same $\bar{a}$ as Ref.~\cite{Nandi:2018mzm}. 
We compare $\bar{b}_{[74]}$ obtained in Ref.~\cite{Nandi:2018mzm} and $\bar{b}$ obtained from Eq.~(\ref{eq:b_our}).}
\begin{center}
\begin{tabular}{ c c c c } \hline
$\lambda$ &$\bar{a}$  &$\bar{b}_{[74]}$  &$\bar{b}$ \\ \hline 
$0$         &$1.0000$    &$-0.4002$ &$-0.4002$\\ 
$0.001$     &$1.0000$    &$-0.4002$ &$-0.4002$\\       
$0.01$      &$1.0001$    &$-0.4008$ &$-0.4004$\\       
$0.02$      &$1.0004$    &$-0.4028$ &$-0.4007$\\       
$0.03$      &$1.0009$    &$-0.4046$ &$-0.4014$\\       
$0.04$      &$1.0016$    &$-0.4105$ &$-0.4022$\\       
$0.05$      &$1.0025$    &$-0.4163$ &$-0.4034$\\ \hline  
\end{tabular}
\end{center}
\end{table}

The deflection angle was calculated also by Ovgun~\cite{Ovgun:2018fnk}, in the coordinates $(t, r, \varphi, \vartheta)$, 
\begin{eqnarray}\label{eq:Ovgun} 
\bar{b}=\frac{\log(6)\sqrt{-2\lambda^2+1}}{\log(10)2\lambda^2-1} +r_{\mathrm{m}} -\pi.
\end{eqnarray}
See Eq.~(2.32) in Ref.~\cite{Ovgun:2018fnk}.
Bhattacharya and Karimov pointed out that $\bar{b}$ obtained in Ref.~\cite{Ovgun:2018fnk}, i.e., Eq.~(\ref{eq:Ovgun}), is in error~\cite{Bhattacharya:2018leh}.

\subsection{$\lambda>\sqrt{2}/2$}
We consider the case of $\lambda>\sqrt{2}/2$. 
In this case, the throat is the photon sphere at $r=r_{\mathrm{m}}=r_{\mathrm{th}}=2M(1+\lambda^2)$
and the critical impact parameter is $b_{\mathrm{m}}=2M(1+\lambda^2)^\frac{3}{2}/\lambda$.
We notice that 
the regular factor $R(z,r_{\mathrm{m}})$ is given by
\begin{eqnarray}
R(z, r_{\mathrm{m}})=2\sqrt{\frac{\lambda^2+z}{(1+\lambda^2)z}}
\end{eqnarray}
and the form of $z^{-\frac{1}{2}}$ is not suitable for analysis of the strong deflection limit. 
Thus, we express $I(r_0)$ as
\begin{eqnarray}
I(r_0)= \int^{1}_{0} S(z,r_0)h(z,r_0)dz,
\end{eqnarray}
where a new regular factor $S(z,r_0)$ and a new divergent factor $h(z,r_0)$ are given by
\begin{eqnarray}
S(z,r_0)\equiv 2\sqrt{r_0-2M(1-z)}
\end{eqnarray}
and
\begin{eqnarray}
&&h(z,r_0)\equiv \nonumber\\
&&\sqrt{\frac{1}{\gamma_0 z+\eta_0 z^2 + \left[ -\frac{2M(r_0-r_{\mathrm{m}})}{r_0}+r_{\mathrm{m}} \beta_0 \right]z^3-\frac{2Mr_{\mathrm{m}}}{r_0}z^4 }}, \nonumber\\
\end{eqnarray}
where $\gamma_0$ and $\eta_0$ are defined as
\begin{eqnarray}
&&\gamma_0=\gamma (r_0) \equiv  (r_0-r_{\mathrm{m}})\alpha_0, \\
&&\eta_0=\eta (r_0) \equiv (r_0-r_{\mathrm{m}})\beta_0+r_{\mathrm{m}} \alpha_0.
\end{eqnarray}
Notice that $\alpha_{\mathrm{m}}=(-1+2\lambda^2)/(1+\lambda^2)$ and $\beta_{\mathrm{m}}=(2-\lambda^2)/(1+\lambda^2)$ in the case of $\lambda>\sqrt{2}/2$.
Since we obtain $\gamma_{\mathrm{m}}=0$ and $\eta_{\mathrm{m}}=2M(2\lambda^2-1)$,
the integral of 
$h_0(z,r_0)$ defined as
\begin{eqnarray}
h_0(z,r_0) \equiv \sqrt{\frac{1}{\gamma_0 z+\eta_0 z^2 }}
\end{eqnarray}
gives the divergent part of $I(r_0)$ in the strong deflection limit $r_0 \rightarrow r_{\mathrm{m}}$.
We separate $I(r_0)$ into the divergent part $I_{\mathrm{d}}(r_0)$ and a regular part $I_{\mathrm{r}}(r_0)$, i.e.,
\begin{eqnarray}
I(r_0)=I_{\mathrm{d}}(r_0)+I_{\mathrm{r}}(r_0),
\end{eqnarray}
where 
\begin{eqnarray}
&&I_{\mathrm{d}}(r_0)=\int^{1}_{0} S(0,r_{\mathrm{m}})h_0(z,r_0) dz, \\
&&I_{\mathrm{r}}(r_0)=\int^{1}_{0} k(z,r_0) dz.
\end{eqnarray}
Here, $k(z,r_0)$ is 
\begin{eqnarray}
k(z,r_0)\equiv S(z,r_0)h(z,r_0)-S(0,r_{\mathrm{m}})h_0(z,r_0).
\end{eqnarray}

The divergent part $I_{\mathrm{d}}(r_0)$ can be integrated and we obtain
\begin{eqnarray}\label{eq:I_d_0} 
&&I_{\mathrm{d}}(r_0)=  \frac{2S(0,r_{\mathrm{m}})}{\sqrt{\eta_0}} \log \frac{\sqrt{\eta_0}+\sqrt{\gamma_0+\eta_0}}{\sqrt{\gamma_0}}.
\end{eqnarray}
By substituting $\gamma_0$ and $\eta_0$ expanded in powers of $(r_0-r_{\mathrm{m}})$ as
\begin{eqnarray}
&&\gamma_0=\alpha_{\mathrm{m}} (r_0-r_{\mathrm{m}}) +O\left((r_0-r_{\mathrm{m}})^2\right),\\
&&\eta_0=2M(2\lambda^2-1)+ O \left( r_0-r_{\mathrm{m}} \right)
\end{eqnarray}
into Eq.~(\ref{eq:I_d_0}), we obtain
\begin{eqnarray}
I_{\mathrm{d}}(r_0)&=&-\frac{2\lambda }{\sqrt{2\lambda^2-1}} \log \left( \frac{r_0}{r_{\mathrm{m}}}-1 \right) +\frac{4\lambda \log 2}{\sqrt{2\lambda^2-1}} \nonumber\\
&&+O\left( \left( \frac{r_0}{r_{\mathrm{m}}}-1 \right) \log \left( \frac{r_0}{r_{\mathrm{m}}}-1 \right)\right). 
\end{eqnarray}
From the impact parameter~$b$ which is expanded in powers of~$(r_0-r_{\mathrm{m}})$ as
\begin{eqnarray}
b=b_{\mathrm{m}}+\frac{(2\lambda^2-1)\sqrt{1+\lambda^2}}{2\lambda^3}(r_0-r_{\mathrm{m}}) +O\left( (r_0-r_{\mathrm{m}})^2 \right),\nonumber\\
\end{eqnarray}
we obtain $I_{\mathrm{d}}=I_{\mathrm{d}}(b)$ as
\begin{eqnarray}
I_{\mathrm{d}}(b)&=&-\frac{2\lambda}{\sqrt{2\lambda^2-1}} \log \left( \frac{b}{b_{\mathrm{m}}}-1 \right) \nonumber\\
&&+\frac{2 \lambda}{\sqrt{2\lambda^2-1}} \log \frac{2(2 \lambda^2-1)}{\lambda^2} \nonumber\\
&&+O\left( \left( \frac{b}{b_{\mathrm{m}}}-1 \right) \log \left( \frac{b}{b_{\mathrm{m}}}-1 \right)\right). 
\end{eqnarray}

The regular part~$I_{\mathrm{r}}(r_0)$ can be expanded in powers of~$(r_0-r_{\mathrm{m}})$ as
\begin{eqnarray}
I_{\mathrm{r}}(r_0)=\sum^{\infty}_{j=0} \frac{1}{j!} (r_0-r_{\mathrm{m}})^j \int^1_0 \left. \frac{\partial^j k}{\partial r^j_0} \right|_{r_0=r_{\mathrm{m}}} dz
\end{eqnarray}
and the term of $j=0$ gives 
\begin{eqnarray}
I_{\mathrm{r}}(r_0)
&=&\int^1_0 k(z, r_{\mathrm{m}})dz \nonumber\\
&&+O\left( \left( \frac{r_0}{r_{\mathrm{m}}}-1 \right) \log \left( \frac{r_0}{r_{\mathrm{m}}}-1 \right)\right), 
\end{eqnarray}
where $k(z, r_{\mathrm{m}})$ is given by
\begin{eqnarray}
k(z, r_{\mathrm{m}})
=2\sqrt{\frac{\lambda^2+z}{(2\lambda^2-1)z^2+(2-\lambda^2)z^3-z^4}}-\frac{2\lambda}{z\sqrt{2\lambda^2-1}}.\nonumber\\
\end{eqnarray}
Therefore, we have obtained   
\begin{eqnarray}
\bar{a}&=&\frac{2\lambda}{\sqrt{2\lambda^2-1}},\\
\bar{b}&=&\frac{2 \lambda}{\sqrt{2\lambda^2-1}} \log \frac{2(2 \lambda^2-1)}{\lambda^2} +I_{\mathrm{r}} -\pi.
\end{eqnarray}

\subsection{$\lambda=\sqrt{2}/2$}
In the case of $\lambda=\sqrt{2}/2$, the throat corresponds with the photon sphere, 
i.e., $r_{\mathrm{m}}=r_{\mathrm{th}}=3M$.
The critical impact parameter $b_{\mathrm{m}}$ is given by $b_{\mathrm{m}}=3\sqrt{3}M$.
From $\alpha_{\mathrm{m}}=\gamma_{\mathrm{m}}=\eta_{\mathrm{m}}=0$ and $\beta_{\mathrm{m}}=1$, 
when $r_0=r_\mathrm{m}$,
the divergent factor $h(z,r_0)$ gives
\begin{eqnarray}
h(z,r_{\mathrm{m}}) = \sqrt{\frac{1}{3Mz^3-2Mz^4}}
\end{eqnarray}
and it causes the integral $I$ to diverge as
\begin{eqnarray}
I\sim \left.z^{-\frac{1}{2}} \right|_{z=0}.
\end{eqnarray}
This implies that $I(r_0)$ has the following form, in the strong deflection limit $r_0\rightarrow r_{\mathrm{m}}=3M$,  
\begin{eqnarray}
I(r_0) = \frac{\mathcal{A}}{\sqrt{\frac{r_0}{r_{\mathrm{m}}}-1}}+\mathcal{B}+O\left(\sqrt{\frac{r_0}{r_{\mathrm{m}}}-1}\right),
\end{eqnarray}
where $\mathcal{A}$ and $\mathcal{B}$ are constant.

We separate the integral $I$ as
\begin{eqnarray}
I=I_{\mathcal{D}}+I_{\mathcal{R}},
\end{eqnarray}
where a divergent part $I_{\mathcal{D}}$ and a regular part $I_{\mathcal{R}}$ are defined by
\begin{eqnarray}
&&I_{\mathcal{D}} \equiv \int^{1}_{0} S(0,r_{\mathrm{m}})h(z,r_{\mathrm{m}}) dz, \\
&&I_{\mathcal{R}} \equiv \int^{1}_{0} q(z,r_0) dz,
\end{eqnarray}
respectively, and where
\begin{eqnarray}
q(z,r_0)\equiv S(z,r_0)h(z,r_0)-S(0,r_{\mathrm{m}})h(z,r_{\mathrm{m}}).
\end{eqnarray}

The divergent part $I_\mathcal{D}$ can be integrated as
\begin{eqnarray}\label{eq:I_mD} 
I_{\mathcal{D}}
&\sim&\left. \frac{4}{\sqrt{3z}} \right|_{z=0} -\frac{4}{3} \nonumber\\
&=&\frac{4\sqrt{3}}{3\sqrt{\frac{r_0}{r_{\mathrm{m}}}-1}} -\frac{4}{3}+O\left(\sqrt{\frac{r_0}{r_{\mathrm{m}}}-1}\right).
\end{eqnarray}
From Eq.~(\ref{eq:b_r}), 
$I_\mathcal{D}=I_\mathcal{D}(b)$ is given by
\begin{eqnarray}
I_\mathcal{D}(b)
=\frac{2^\frac{7}{4} 3^{-\frac{1}{4}}}{\left( \frac{b}{b_{\mathrm{m}}}-1 \right)^\frac{1}{4}} -\frac{4}{3}  
+O\left( \left( \frac{b}{b_{\mathrm{m}}}-1 \right)^\frac{3}{4} \right).
\end{eqnarray}

The regular part $I_\mathcal{R}(r_0)$ can be expanded in powers of $(r_0-r_{\mathrm{m}})$ as
\begin{eqnarray}
I_\mathcal{R}(r_0)=\sum^{\infty}_{j=0} \frac{1}{j!} (r_0-r_{\mathrm{m}})^j \int^1_0 \left. \frac{\partial^j q}{\partial r^j_0} \right|_{r_0=r_{\mathrm{m}}} dz
\end{eqnarray}
and the term of $j=0$ gives 
\begin{eqnarray}
I_\mathcal{R}(r_0)
=\int^1_0 q(z, r_{\mathrm{m}})dz 
+O\left(\sqrt{\frac{r_0}{r_{\mathrm{m}}}-1}\right), 
\end{eqnarray}
where $q(z, r_{\mathrm{m}})$ is given by
\begin{eqnarray}
q(z, r_{\mathrm{m}})
=\frac{2\left( \sqrt{1+2z}-1 \right)}{z\sqrt{3z-2z^2}}.
\end{eqnarray}
We obtain $I_\mathcal{R}(r_{\mathrm{m}})=2.3671$ in a numerical calculation. 
Therefore, the deflection angle of the light in the strong deflection limit is obtained as
\begin{eqnarray}
\alpha_{\mathrm{def}}(b)
=\frac{\bar{c}}{\left( \frac{b}{b_{\mathrm{m}}}-1 \right)^\frac{1}{4}} +\bar{d}
+O\left( \left( \frac{b}{b_{\mathrm{m}}}-1 \right)^\frac{3}{4} \right),
\end{eqnarray}
where $\bar{c}$ and $\bar{d}$ are given by
\begin{eqnarray}
&&\bar{c}\equiv 2^\frac{7}{4} 3^{-\frac{1}{4}}= 2.5558 \nonumber\\
&&\bar{d}\equiv -\frac{4}{3}+I_\mathcal{R}(r_{\mathrm{m}})-\pi=-2.1078.
\end{eqnarray}

\section{Observables in the strong deflection limit}
We consider a small angle lens equation~\cite{Bozza:2008ev}
\begin{eqnarray}\label{eq:lens}
D_{\mathrm{LS}}\bar{\alpha}=D_{\mathrm{OS}}(\theta-\phi),
\end{eqnarray}
where $D_{\mathrm{LS}}$ and $D_{\mathrm{OS}}$ are angular distances between a lens object and a source object and between the observer and the source object, respectively,
$\bar{\alpha}$ is an effective deflection angle defined by
\begin{eqnarray}\label{eq:def_bar_alpha}
\bar{\alpha}\equiv \alpha_{\mathrm{def}} \;  \mathrm{mod} \;  2\pi,
\end{eqnarray}
$\theta$ is an image angle, and $\phi$ is a source angle 
as shown Fig.~4.
\begin{figure}[htbp]
\begin{center}
\includegraphics[width=87mm]{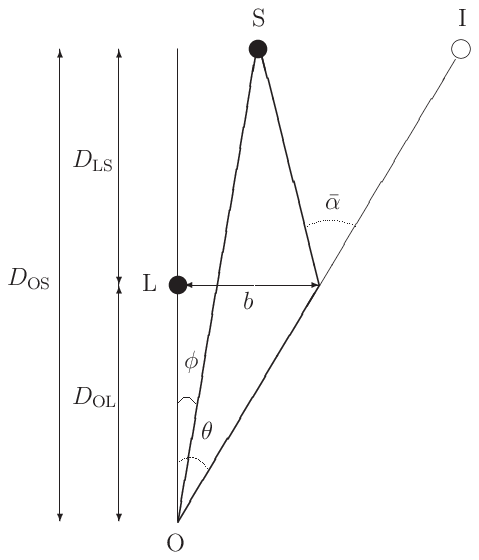}
\end{center}
\caption{
Lens configuration. A light ray, which is emitted by a source S with a source angle $\phi$ and reflected by a lens L with a deflection angle $\alpha_{\mathrm{def}}$ and an impact parameter $b$, reaches an observer~O.
The observer sees an image I with an image angle $\theta$.  
Here, $\bar{\alpha}=\alpha_{\mathrm{def}} -2\pi n $ is the effective deflection angle of the light with a winding number $n$.
$D_{\mathrm{OS}}$, $D_{\mathrm{LS}}$, and $D_{\mathrm{OL}}=D_{\mathrm{OS}}-D_{\mathrm{LS}}$ are distances between the observer and the source, the lens and the source, and the observer and the lens, respectively.
}
\end{figure}
Here, we have assumed that all the angles are small, i.e., $\bar{\alpha}$, $\theta$, $\phi \ll 1$. 
Under the assumption, an impact parameter $b$ is expressed by $b=\theta D_{\mathrm{OL}}$, where 
$D_{\mathrm{OL}}=D_{\mathrm{OS}}-D_{\mathrm{LS}}$ is an angular distance between the observer and the lens.

We can express the deflection angle $\alpha_{\mathrm{def}}$ of a light ray which rotates around the photon sphere $n$ times as 
\begin{eqnarray}\label{eq:alpha_bar_alpha}
\alpha_{\mathrm{def}} = \bar{\alpha} + 2\pi n,
\end{eqnarray}
where the winding number $n$ is a positive integer in this section and $n=0$ in Sec.~V.
We define $\theta^0_n$ as 
\begin{eqnarray}\label{eq:def_theta0n}
\alpha_{\mathrm{def}}(\theta^0_n)= 2 \pi n.
\end{eqnarray}
We expand the deflection angle $\alpha_{\mathrm{def}}(\theta)$ around $\theta=\theta^0_n$ as 
\begin{eqnarray}\label{eq:def_expand}
\alpha_{\mathrm{def}}(\theta)
&=&\alpha_{\mathrm{def}}(\theta^0_n) +\left. \frac{d\alpha_{\mathrm{def}}}{d\theta}\right|_{\theta=\theta^0_n} \left( \theta-\theta^0_n \right) \nonumber\\
&&+O\left( \left( \theta-\theta^0_n \right)^2 \right). 
\end{eqnarray}

\subsection{$\lambda \neq \sqrt{2}/2$}
For $\lambda \neq \sqrt{2}/2$, 
we can rewrite the deflection angle in the strong deflection limit as 
\begin{eqnarray}\label{eq:def_strong3} 
\alpha_{\mathrm{def}}(\theta)&=& -\bar{a} \log \left( \frac{\theta}{\theta_\infty}-1 \right) + \bar{b} \nonumber\\
&&+O\left( \left( \frac{\theta}{\theta_\infty}-1 \right) \log \left( \frac{\theta}{\theta_\infty}-1 \right)\right), 
\end{eqnarray}
where $\theta_\infty \equiv b_{\mathrm{m}}/D_{\mathrm{OL}}$ is the image angle of the photon sphere~\footnote{
It is known that Eq.~(\ref{eq:def_strong3}) is a good approximation in some examples by comparing the exact deflection angle
even if the winding number $n=1$. See Ref.~\cite{Tsukamoto:2014dta} as an example.}. 
In this case, we obtain 
\begin{eqnarray}\label{eq:dalpha_dtheta3}
\left. \frac{d\alpha_{\mathrm{def}}}{d\theta}\right|_{\theta=\theta^0_n}=-\frac{\bar{a}}{\theta^0_n-\theta_{\infty}}
\end{eqnarray}
and, from Eqs.~(\ref{eq:def_theta0n}) and (\ref{eq:def_strong3}),
\begin{eqnarray}\label{eq:theta0n}
\theta^0_n= \left[ 1+ \exp \left( \frac{\bar{b}-2\pi n}{\bar{a}} \right) \right] \theta_\infty.
\end{eqnarray}

We consider the solution $\theta=\theta_n$ of the lens equation~(\ref{eq:lens}) for the light ray with the winding number $n$.
From Eqs.~(\ref{eq:alpha_bar_alpha})-(\ref{eq:def_expand}), (\ref{eq:dalpha_dtheta3}), and (\ref{eq:theta0n}),
we get
\begin{eqnarray}\label{eq:bar_alpha3}
\bar{\alpha}(\theta_n)=\frac{\bar{a}e^{\frac{-\bar{b}+2 \pi n}{\bar{a}}}}{\theta_\infty} \left(  \theta^0_n-\theta_n \right).
\end{eqnarray}
From Eqs.~(\ref{eq:lens}) and (\ref{eq:bar_alpha3}), we obtain the image angle 
\begin{eqnarray}\label{eq:theta_n2}
\theta_n(\phi)\sim \theta^0_n+\frac{\theta_\infty e^{\frac{\bar{b}-2\pi n}{\bar{a}}} D_{\mathrm{OS}} \left( \phi-\theta^0_n \right)}{\bar{a}D_{\mathrm{LS}}}
\end{eqnarray} 
and its Einstein ring angle $\theta_{\mathrm{E} n}$ as
\begin{eqnarray}
\theta_{\mathrm{E} n} \equiv \theta_n(0) = \theta^0_n \left[ 1 -\frac{\theta_\infty e^{\frac{\bar{b}-2\pi n}{\bar{a}}} D_{\mathrm{OS}} }{\bar{a}D_{\mathrm{LS}}} \right].
\end{eqnarray} 
We get the magnification of the image 
\begin{eqnarray}
\mu_n 
&\equiv& \frac{\theta_n}{\phi} \frac{d\theta_n}{d\phi} \nonumber\\
&\sim&  \frac{\theta_\infty^2 D_{\mathrm{OS}} \left( 1+ e^{\frac{\bar{b}-2\pi n}{\bar{a}}} \right) e^{\frac{\bar{b}-2\pi n}{\bar{a}}}}{\bar{a} D_{\mathrm{LS}} \phi},
\end{eqnarray}
the sum of the magnifications of the infinite numbers of images
\begin{eqnarray}
\sum^\infty_{n=1} \mu_n  
\sim  \frac{\theta_\infty^2 D_{\mathrm{OS}} \left( 1+ e^{\frac{2\pi}{\bar{a}}}+e^{\frac{\bar{b}}{\bar{a}}} \right) e^{\frac{\bar{b}}{\bar{a}}}}
{\bar{a} D_{\mathrm{LS}} \phi \left( e^{\frac{4\pi}{\bar{a}}} -1 \right) },
\end{eqnarray}
the magnification without the outermost image
\begin{eqnarray}
\sum^\infty_{n=2} \mu_n  
\sim  \frac{\theta_\infty^2 D_{\mathrm{OS}} \left( e^{\frac{4\pi}{\bar{a}}}+ e^{\frac{2\pi}{\bar{a}}}+e^{\frac{\bar{b}}{\bar{a}}} \right) e^{\frac{\bar{b}-4\pi}{\bar{a}}} }
{\bar{a} D_{\mathrm{LS}} \phi \left( e^{\frac{4\pi}{\bar{a}}} -1 \right) }, 
\end{eqnarray}
and the ratio of the magnification of the outermost image to the others 
\begin{eqnarray}
\mathrm{r}\equiv \frac{\mu_1}{\sum^\infty_{n=2} \mu_n}\sim \frac{\left( e^{\frac{4\pi}{\bar{a}}}-1 \right) \left( e^{\frac{2\pi}{\bar{a}}}+e^{\frac{\bar{b}}{\bar{a}}} \right)}
{ e^{\frac{4\pi}{\bar{a}}}+ e^{\frac{2\pi}{\bar{a}}}+e^{\frac{\bar{b}}{\bar{a}}}}.
\end{eqnarray}
The difference of the image angles between the outermost images and the photon sphere is obtained as
\begin{eqnarray}
\mathrm{s}\equiv \theta_1-\theta_\infty \sim \theta^0_1- \theta^0_\infty= \theta_\infty e^{\frac{\bar{b}-2\pi}{\bar{a}}}.
\end{eqnarray}

\subsection{$\lambda=\sqrt{2}/2$}
For $\lambda=\sqrt{2}/2$, the deflection angle in the strong deflection limit can be expressed by
\begin{eqnarray}\label{eq:def_strong12}
\alpha_{\mathrm{def}}(\theta)
=\frac{\bar{c}}{\left(\frac{\theta}{\theta_\infty}-1\right)^{\frac{1}{4}}}+\bar{d} +O\left(\left(\frac{\theta}{\theta_\infty}-1\right)^{\frac{3}{4}}\right).
\end{eqnarray}
In this case, 
we obtain 
\begin{eqnarray}\label{eq:dalpha_dtheta12}
\left. \frac{d\alpha_{\mathrm{def}}}{d\theta}\right|_{\theta=\theta^0_n}=-\frac{\bar{c}}{4\theta_\infty} \left( \frac{\theta^0_n}{\theta_\infty}-1 \right)^{-\frac{5}{4}}
\end{eqnarray}
and, from Eqs.~(\ref{eq:def_theta0n}) and (\ref{eq:def_strong12}),
\begin{eqnarray}\label{eq:theta0n2}
\theta^0_n= \left[ 1+ \left( \frac{\bar{c}}{2\pi n -\bar{d}} \right)^4 \right] \theta_\infty.
\end{eqnarray}
Thus, from Eqs.~(\ref{eq:alpha_bar_alpha})-(\ref{eq:def_expand}), (\ref{eq:dalpha_dtheta12}), and (\ref{eq:theta0n2}), 
the effective deflection angle is obtained by
\begin{eqnarray}\label{eq:bar_alpha12}
\bar{\alpha}(\theta_n)= \frac{\left(2 \pi n -\bar{d}\right)^5}{4 \theta_\infty \bar{c}^4} \left( \theta^0_n - \theta_n \right).
\end{eqnarray}
From Eqs.~(\ref{eq:lens}) and (\ref{eq:bar_alpha12}), we get the image angle with the winding number~$n$
\begin{eqnarray}
\theta_n(\phi)\sim \theta^0_n+\frac{4\bar{c}^4D_{\mathrm{OS}}\theta_\infty(\phi-\theta^0_n)}{(2\pi n-\bar{d})^5D_{\mathrm{LS}}}
\end{eqnarray}
and the Einstein ring angle  
\begin{eqnarray}
\theta_{\mathrm{E}n} \equiv \theta_n(0)\sim \theta^0_n \left[1 -\frac{4\bar{c}^4D_{\mathrm{OS}}\theta_\infty}{(2\pi n-\bar{d})^5D_{\mathrm{LS}}} \right].
\end{eqnarray}
Therefore, we obtain the magnification of the image
\begin{eqnarray}
\mu_n \equiv \frac{\theta_n}{\phi} \frac{d\theta_n}{d\phi}
\sim \frac{4 \bar{c}^4 D_{\mathrm{OS}}\theta_\infty^2F(n)}{D_{\mathrm{LS}} \phi},
\end{eqnarray}
where $F(n)$ defined by
\begin{eqnarray}
F(n)\equiv \frac{1+ \left( \frac{\bar{c}}{2\pi n -\bar{d}} \right)^4 }{(2\pi n-\bar{d})^5}
\end{eqnarray}
can be calculated numerically. 
We can calculate the magnification of the infinite number of images 
\begin{eqnarray}
\sum^\infty_{n=1} \mu_n  \sim \frac{4 \bar{c}^4 D_{\mathrm{OS}}\theta_\infty^2}{D_{\mathrm{LS}} \phi} \sum^\infty_{n=1}F(n) 
\end{eqnarray}
and the sum of the magnifications of the images without the outermost image 
\begin{eqnarray}
\sum^\infty_{n=2} \mu_n \sim \frac{4 \bar{c}^4 D_{\mathrm{OS}}\theta_\infty^2}{D_{\mathrm{LS}} \phi} \sum^\infty_{n=2}F(n),
\end{eqnarray}
where 
\begin{eqnarray}
&&\sum^\infty_{n=1} F(n) \sim 2.6077 \times 10^{-5}, \\
&&\sum^\infty_{n=2} F(n) \sim 1.8303 \times 10^{-6}.
\end{eqnarray}
Therefore, the ratio of the magnifications of the outermost image to the other images is obtained as
\begin{eqnarray}
\mathrm{r}= \frac{\mu_1}{\sum^\infty_{n=2} \mu_n}\sim \frac{F(1)}{\sum^\infty_{n=2} F(n)}=13.248,
\end{eqnarray}
where we have used
\begin{eqnarray}
F(1)\sim 2.4247 \times 10^{-5}. 
\end{eqnarray}
The difference of the image angles between the outermost images and the photon sphere is 
\begin{eqnarray}
\mathrm{s}= \theta_1-\theta_\infty \sim \theta^0_1- \theta^0_\infty= \left( \frac{\bar{c}}{2\pi -\bar{d}} \right)^4 \theta_\infty .
\end{eqnarray}

\section{Gravitational lens under the weak-field approximation}
Under a weak-field approximation $M/r_0 \ll 1$,
the line element is given by
\begin{eqnarray}\label{eq:line_element0} 
ds^2
&=&-\left( 1-\frac{2M}{r} \right)dt^2 +\left[ 1+\frac{2M\left( 1+\lambda^2 \right)}{r} \right]dr^2 \nonumber\\
&&+r^2 \left( d \vartheta^2 +\sin^2\vartheta d \varphi^2 \right).
\end{eqnarray}

From Eq.~$(8.5.8)$ in Ref.~\cite{Weinberg:1972kfs},
the deflection angle $\alpha_{\mathrm{def}}$ of a light ray is obtained as
\begin{eqnarray}\label{eq:deflection_angle_weak} 
\alpha_{\mathrm{def}}
=\frac{4M_*}{r_0} + O\left(\left( \frac{M_*}{r_0} \right)^2\right)
\end{eqnarray}
where $M_*$ is 
\begin{eqnarray}
M_*\equiv M\left( 1+ \frac{\lambda^2}{2} \right).
\end{eqnarray}
From Eq.~(\ref{eq:b_0}), $b/r_0=1+O(M/r_0)$ is satisfied. 
Thus, the deflection angle is rewritten as 
\begin{eqnarray}\label{eq:deflection_angle_weak2} 
\alpha_{\mathrm{def}}
=\frac{4M_*}{b} + O\left(\left( \frac{M_*}{b} \right)^2\right).
\end{eqnarray}
This is the same as Eq.~$(2.21)$ in Ref.~\cite{Ovgun:2018fnk}.

From Eqs.~(\ref{eq:alpha_bar_alpha}), $(\ref{eq:deflection_angle_weak2})$, $n=0$, and $b=\theta D_{\mathrm{OL}}$,
the solution $\theta=\theta_{\pm0}(\phi)$ of the lens equation (\ref{eq:lens}) is given by
\begin{eqnarray}
\theta_{\pm0}(\phi)=\frac{1}{2}\left( \phi \pm \sqrt{ \phi^2+  4\theta_{\mathrm{E}0}^2  } \right),
\end{eqnarray}
where $\theta_{\mathrm{E}0}$ is the angle of an Einstein ring defined as
\begin{eqnarray}
\theta_{\mathrm{E}0}\equiv \theta_{+0}(0)= \sqrt{ \frac{4M_*D_{\mathrm{LS}}}{D_{\mathrm{OS}}D_{\mathrm{OL}}} }.
\end{eqnarray}
Notice that $\theta_{-0}(\phi)$ has a negative value and its impact parameter is also negative.
We obtain the magnifications of the images 
\begin{eqnarray}
\mu_{\pm0}
&\equiv& \frac{\theta_{\pm0}}{\phi} \frac{d\theta_{\pm0}}{d \phi} \nonumber\\
&=& \frac{1}{4} \left( 2 \pm \frac{\phi}{\sqrt{\phi^2+4\theta_{\mathrm{E}0}^2}} \pm \frac{\sqrt{\phi^2+4\theta_{\mathrm{E}0}^2}}{\phi} \right)
\end{eqnarray}
and its total magnification 
\begin{eqnarray}
\mu_{0\mathrm{tot}}
&\equiv& \left| \mu_{+0} \right|+\left| \mu_{-0} \right|  \nonumber\\
&=& \frac{1}{2} \left( \frac{\phi}{\sqrt{\phi^2+4\theta_{\mathrm{E}0}^2}} + \frac{\sqrt{\phi^2+4\theta_{\mathrm{E}0}^2}}{\phi} \right).
\end{eqnarray}

\section{Discussion and conclusion}
From Secs.~I to IV, we have concentrated on an infinite number of images with positive impact parameters $b$ or positive image angles $\theta_n$. 
The each image has a partner with a negative impact parameter.
The image angle $\theta_{-n}$ and the magnification $\mu_{-n}$ of the partner of the image with $\theta_n$ are given by $\theta_{-n} \sim -\theta_{n}$ and $\mu_{-n} \sim -\mu_{n}$, respectively.
Thus, the diameter of the pair images on a sky is obtained as $\theta_n-\theta_{-n} \sim 2\theta_{n}$ 
and the total magnification of the pair images is given by $\mu_{n\mathrm{tot}}\equiv \left| \mu_{n} \right|+\left| \mu_{-n} \right| \sim 2 \left| \mu_{n} \right|$.  
The observables and the parameters $\bar{a}$, $\bar{b}$, $\bar{c}$, and $\bar{d}$ of the deflection angle in the strong deflection limit are summarized in Table~II.  
\begin{table*}[htbp]
 \label{table:II}
 \caption{Parameters $\bar{a}$, $\bar{b}$, $\bar{c}$, and $\bar{d}$ in the deflection angle in the strong deflection limit, the diameters of the innermost ring~$2\theta_{\infty}$, 
 and the outermost ring~$2\theta_{\mathrm{E}1}$ among rings scattered by the photon sphere, the difference of the radii of the outermost ring and the innermost ring $\mathrm{s}=\theta_{\mathrm{E}1}-\theta_\infty$, 
 the magnification of the pair images of the outermost ring $\mu_{1\mathrm{tot}} \sim 2 \left| \mu_{1} \right|$, the ratio of the magnification of the outermost ring to the other rings $\mathrm{r}= \mu_1/\sum^\infty_{n=2} \mu_n$ for given $\lambda$.
 Here we have set $D_{\mathrm{OS}}=16$kpc, $D_{\mathrm{OL}}=8$kpc, $D_{\mathrm{LS}}=D_{\mathrm{OS}}-D_{\mathrm{OL}}=8$kpc, $M_*=M\left( 1+ \lambda^2/2 \right)=4\times 10^6 M_{\odot}$, 
 and the source angle $\phi=1$ arcsecond for $\mu_{1\mathrm{tot}}$.
 Notice that the diameter of the Einstein ring $2\theta_{\mathrm{E}0}=2.8618$~arcsecond 
 and the magnification of a pair of images $\mu_{0\mathrm{tot}}=1.6807$ for the source angle $\phi=1$ arcsecond do not depend on $\lambda$ under the weak-field approximation 
 since we make $M_*$ constant. 
 We find that $2\theta_{\infty}$ monotonically decreases as $\lambda$ increases from $0$ to $\sqrt{2}$ and it monotonically increases as $\lambda$ increases from $\sqrt{2}$ to $\infty$ if $M_*$ is constant.}
\begin{center}
\begin{tabular}{ c c c c c c c c c c c} \hline
$\lambda$          	              &$0$         &$0.4$        &$0.6$      &$0.7$       &$\sqrt{2}/2$ &$0.71$       &$0.8$     &$1$        &$\sqrt{2}$  &$5$ \\ \hline
$\bar{a}$        	              &$1.0000$    &$1.2127$     &$1.8898$   &$7.0711$    &$--$         &$15.681$     &$3.0237$  &$2.000$    &$1.6330$    &$1.4286$ \\ 
$\bar{b}$        	              &$-0.4002$   &$-0.7591$    &$-2.8784$  &$-40.837$   &$--$         &$-116.02$    &$-4.8720$ &$-1.3632$  &$-0.77941$  &$-0.7264$ \\ 
$\bar{c}$    	                      &$--$        &$--$         &$--$       &$--$        &$2.5558$     &$--$         &$--$      &$--$       &$--$        &$--$ \\ 
$\bar{d}$                             &$--$        &$--$         &$--$       &$--$        &$-2.1078$    &$--$         &$--$      &$--$       &$--$        &$--$ \\ 
$2\theta_{\infty}$~[$\mu$as]          &$51.580$    &$47.759$     &$43.712$   &$41.430$    &$41.264$     &$41.197$     &$39.485$  &$37.436$   &$36.473$    &$38.993$ \\ 
$2\theta_{\mathrm{E}1}$~[$\mu$as]     &$51.645$    &$47.903$     &$44.055$   &$41.483$    &$41.619$     &$41.214$     &$40.472$  &$38.254$   &$36.955$    &$39.282$ \\ 
$\mathrm{s}$~[$\mu$as]                &$0.032277$  &$0.071785$   &$0.17145$  &$0.026439$  &$0.17758$    &$0.0084446$  &$0.49338$ &$0.40913$  &$0.24135$   &$0.14423$ \\ 
$\mu_{1\mathrm{tot}} \times 10^{17}$  &$1.6163$    &$2.7495$     &$3.8755$   &$0.15039$   &$3.4162$     &$0.021521$   &$6.4033$  &$7.5878$   &$5.2960$    &$3.8453$ \\ 
$\mathrm{r}$     	              &$535.16$    &$177.42$     &$26.996$   &$1.4330$    &$13.248$     &$0.49297$    &$7.1431$  &$22.604$   &$46.476$    &$80.886$ \\ 
\hline
\end{tabular}
\end{center}
\end{table*}

As shown in Sec~V, the gravitational lensing under the weak-field approximation is not characterized by $M$ but $M_*$.
Under an assumption that $M_*$ is constant, 
the size of the photon sphere $\theta_{\infty}$ monotonically decreases as $\lambda$ increases from $0$ to $\sqrt{2}$ and $\theta_{\infty}$ monotonically increases as $\lambda$ increases from $\sqrt{2}$ to $\infty$.
The minimal value of $\theta_\infty$ is given by $3\sqrt{6}M_*/(4D_{\mathrm{OL}})$ for $\lambda=\sqrt{2}$.~\footnote{When $M$ is constant, 
the size of the photon sphere $\theta_\infty$ monotonically increases as $\lambda$ increases from 0 to $\infty$ and it takes a constant and minimum value $3\sqrt{3}M/D_{\mathrm{OL}}$ for $\lambda\leq \sqrt{2}/2$.}

We summarize our result. 
We have shown that the Damour-Solodukhin wormhole with two flat regions has two photon spheres and an antiphoton sphere for $\lambda<\sqrt{2}/2$ 
and only one photon sphere for $\lambda \geq \sqrt{2}/2$ and the photon sphere is marginally unstable when $\lambda = \sqrt{2}/2$.
We have reexamined that deflection angle in the strong deflection limit for $\lambda<\sqrt{2}/2$ and we have extent the analysis for $\lambda=\sqrt{2}/2$ and $\lambda> \sqrt{2}/2$.
We have found that the deflection angle of a light ray reflected by the marginally unstable photon sphere diverges nonlogarithmically in the strong deflection limit for $\lambda = \sqrt{2}/2$, 
while the deflection angle of the light reflected by the photon sphere diverges logarithmically for $\lambda \neq \sqrt{2}/2$.
We expect that our method can be applied for gravitational lenses by marginally unstable photon spheres of various compact objects.

\section*{Acknowledgements}
The author thanks R.~Izmailov, K.~K.~Nandi, A.~\"{O}vg\"{u}n, and an anonymous referee for their useful comments.
\appendix

\section{Arnowitt-Deser-Misner masses}
Wormholes have two ADM masses since the ADM mass is defined in every asymptotically flat region~\cite{Poisson}.
For simplicity, we have assumed the equal ADM masses of the Damour-Solodukhin wormhole.
See Visser~\cite{Visser_1995} for the details of the mass of the wormholes.

We show that $M_{\mathrm{ADM}}= M\left(1+\lambda^2 \right)$ is the ADM mass of the wormhole.
By using a radial coordinate $r_*$, which is given by 
\begin{eqnarray}
r_* \equiv \frac{1}{2} \left[ r-M \left( 1+\lambda^2 \right) + \sqrt{r^2- 2M\left( 1+\lambda^2 \right)r}  \right]
\end{eqnarray}
or 
\begin{eqnarray}
r=r_* \left[ 1+ \frac{M\left( 1+\lambda^2 \right)}{2 r_*} \right]^2,
\end{eqnarray}
the line element~$(\ref{eq:line_element0})$ under the weak field approximation 
is rewritten as
\begin{eqnarray}\label{eq:line_element_weak} 
ds^2
&=&-\left( 1-\frac{2M}{r(r_*)} \right)dt^2 \nonumber\\
&&+\left[ 1+\frac{2M\left( 1+\lambda^2 \right)}{r_*} \right] \left[ dr_*^2 
+r_*^2 \left( d \vartheta^2 +\sin^2\vartheta d \varphi^2 \right) \right]. \nonumber\\
\end{eqnarray}
We consider the hypersurfaces $\Sigma_{\mathrm{t}}$, which are surfaces of constant $t$ with a unit normal $n_{\alpha}=-(1-M/r)\partial_{\alpha}t$.
The induced metric on $\Sigma_{\mathrm{t}}$ is obtained as
\begin{eqnarray}
h_{ab}dy^{a}dy^{b}
=\left[ 1+\frac{2M\left( 1+\lambda^2 \right)}{r_*} \right] \left[ dr_*^2 
+r_*^2 \left( d \vartheta^2 +\sin^2\vartheta d \varphi^2 \right) \right]. \nonumber\\
\end{eqnarray}
The induced metric on a two-sphere $S_{\mathrm{t}}$ at $r_*=R_{\mathrm{t}}$ 
with a unit normal $r_{a}=\left[1+M(1+\lambda^2)/r_* \right] \partial_{a}r_*$ is
\begin{equation}
\sigma_{AB}d\theta^{A}d\theta^{B}=\left[ 1+\frac{2M\left( 1+\lambda^2 \right)}{R_{\mathrm{t}}} \right]R_{\mathrm{t}}^{2}\left( d \vartheta^2 +\sin^2\vartheta d \varphi^2 \right).
\end{equation}
The extrinsic curvature of $S_{\mathrm{t}}$ embedded in $\Sigma_{\mathrm{t}}$ is obtained as
$k=\sigma^{AB}k_{AB}=r^{a}_{\;\;  \left| a \right.}=2\left[R_{\mathrm{t}}-2M\left( 1+\lambda^2 \right)\right]/R_{\mathrm{t}}^2$,
where $\,_{\left| a \right.}$ is the covariant differentiation on $\Sigma_{\mathrm{t}}$.
The extrinsic curvature of $S_{\mathrm{t}}$ embedded in flat space is given by $k_{0}=2\left[R_{\mathrm{t}}-M\left( 1+\lambda^2 \right)\right]/R_{\mathrm{t}}^2$. 
The ADM mass is obtained as
\begin{eqnarray}
M_{\mathrm{ADM}}
&\equiv&-\frac{1}{8\pi} \lim_{S_{\mathrm{t}}\rightarrow \pm \infty} \oint_{S_{\mathrm{t}}}(k-k_{0})\sqrt{\sigma}d^{2}\theta \nonumber\\
&=&M(1+\lambda^2),
\end{eqnarray}
where $\sigma=R_{\mathrm{t}}^4 \sin^2 \vartheta$.
Therefore, the ADM mass is not $M$ but $M(1+\lambda^2)$.

\section{Violation of energy conditions}
The Ricci tensor and Ricci scalar are given by 
\begin{eqnarray}
&&\mathcal{R}_{tt}=-\frac{M^2 \lambda^2}{r^3 (2 M-r)},\\
&&\mathcal{R}_{rr}=-\frac{(2 r-3 M) M \lambda^2 }{r (r-2 M)^2 \left[r-2M \left( 1+\lambda^2 \right) \right]},\\
&&\mathcal{R}_{\vartheta \vartheta}=-\frac{\lambda^2 M}{2 M-r},\\
&&\mathcal{R}_{\varphi \varphi}=\sin^2 \vartheta \mathcal{R}_{\vartheta \vartheta}.
\end{eqnarray}
and
\begin{eqnarray}
\mathcal{R}=-\frac{2M^2 \lambda^2}{r^2 (r-2 M)^2},
\end{eqnarray}
respectively.
The Einstein tensor is given by 
\begin{eqnarray}
&&G^{t}_{\;\: t}=0,\\
&&G^{r}_{\;\: r}
=-\frac{2M \lambda^2 }{r^2 (r-2M)},\\
&&G^{\vartheta}_{\;\: \vartheta}=G^{\varphi}_{\;\: \varphi}=  \frac{ M (r-M) \lambda^2}{r^2 (r-2 M)^2}.
\end{eqnarray}

The nonzero components of the stress energy tensor $\mathcal{T}^{\mu}_{\;\: \nu}$ are 
$\mathcal{T}^{t}_{\;\:t}=-\rho$, $\mathcal{T}^{r}_{\;\:r}
=p$, and $\mathcal{T}^{\vartheta}_{\;\:\vartheta}=\mathcal{T}^{\varphi}_{\;\:\varphi}=p_{\mathrm{T}}$,
where $\rho$ is the energy density, $p$ is the radial pressure, and $p_{\mathrm{T}}$ is the tangential pressure.

From the Einstein equation $G^{\mu}_{\;\: \nu}=8\pi \mathcal{T}^{\mu}_{\;\: \nu}$, we obtain 
\begin{eqnarray}\label{eq:rho} 
&&\rho=0,\\\label{eq:p} 
&&p=-\frac{M \lambda^2 }{4\pi r^2 (r-2M)},\\\label{eq:pT} 
&&p_{\mathrm{T}}=\frac{ M (r-M) \lambda^2}{8\pi r^2 (r-2 M)^2}.
\end{eqnarray}
The weak, null, and strong energy conditions~\cite{Visser_1995} are violated everywhere because of $\rho+p<0$.


\begin{thebibliography}{99}



\bibitem{Abbott:2016blz}
  B.~P.~Abbott {\it et al.} [LIGO Scientific and Virgo Collaborations],
  Phys.\ Rev.\ Lett.\  {\bf 116}, 061102 (2016).

\bibitem{LIGOScientific:2018mvr} 
  B.~P.~Abbott {\it et al.} [LIGO Scientific and Virgo Collaborations],
  Phys.\ Rev.\ X {\bf 9}, 031040 (2019).
  
\bibitem{Akiyama:2019cqa} 
  K.~Akiyama {\it et al.} [Event Horizon Telescope Collaboration],
  Astrophys.\ J.\  {\bf 875}, L1 (2019).

\bibitem{Claudel:2000yi} 
  C.~M.~Claudel, K.~S.~Virbhadra and G.~F.~R.~Ellis,
  J.\ Math.\ Phys.\  {\bf 42}, 818 (2001).

\bibitem{Perlick_2004_Living_Rev}
V. Perlick,
Living Rev. Relativity {\bf7}, 9 (2004).

\bibitem{Hod:2017xkz} 
  S.~Hod,
  Phys.\ Lett.\ B {\bf 727}, 345 (2013).

\bibitem{Hagihara_1931} 
Y.~Hagihara, 
Jpn.\ J.\ Astron.\ Geophys., {\bf 8}, 67 (1931).

\bibitem{Darwin_1959}
C. Darwin,
Proc. R. Soc. Lond. A {\bf 249}, 180 (1959).

\bibitem{Atkinson_1965}
R.~d'~E. Atkinson,
Astron. J. {\bf 70}, 517 (1965).

\bibitem{Luminet_1979}
J.-P. Luminet,  Astron. Astrophys. {\bf 75}, 228 (1979).

\bibitem{Ohanian_1987}
H. C. Ohanian, Am. J. Phys. {\bf 55}, 428 (1987).

\bibitem{Nemiroff_1993}
R. J. Nemiroff,  Am. J. Phys. {\bf 61}, 619 (1993).

\bibitem{Frittelli_Kling_Newman_2000}
S. Frittelli, T. P. Kling, and E. T. Newman,
Phys. Rev. D {\bf 61}, 064021 (2000).

\bibitem{Virbhadra_Ellis_2000}
K. S. Virbhadra and G. F. R. Ellis,
Phys. Rev. D {\bf 62}, 084003 (2000).

\bibitem{Bozza_Capozziello_Iovane_Scarpetta_2001}
V. Bozza, S. Capozziello, G. Iovane, and G. Scarpetta,
Gen. Relativ. Gravit. {\bf 33}, 1535 (2001).

\bibitem{Bozza:2002zj} 
  V.~Bozza,
  Phys.\ Rev.\ D {\bf 66}, 103001 (2002).

\bibitem{Perlick_2004_Phys_Rev_D} 
V. Perlick, 
Phys. Rev. D {\bf 69}, 064017 (2004). 

\bibitem{Bozza_2010}
V. Bozza,
Gen. Relativ. Gravit. {\bf 42}, 2269 (2010).

\bibitem{Ames_1968} 
W. L. Ames and K. S. Thorne, 
Astrophys.\ J. {\bf 151}, 659 (1968).

\bibitem{Synge:1966okc} 
  J.~L.~Synge,
  Mon.\ Not.\ Roy.\ Astron.\ Soc.\  {\bf 131}, no. 3, 463 (1966).

\bibitem{Yoshino:2019qsh} 
  H.~Yoshino, K.~Takahashi, and K.~i.~Nakao,
  Phys.\ Rev.\ D {\bf 100}, 084062 (2019).

\bibitem{Sanchez:1977si} 
  N.~G.~Sanchez,
  Phys.\ Rev.\ D {\bf 18}, 1030 (1978).

\bibitem{Decanini:2010fz} 
  Y.~Decanini, A.~Folacci, and B.~Raffaelli,
  Phys.\ Rev.\ D {\bf 81}, 104039 (2010).

  \bibitem{Press:1971wr} 
  W.~H.~Press,
  Astrophys.\ J.\  {\bf 170}, L105 (1971).

\bibitem{Goebel_1972} 
 C. J. Goebel, 
Astrophys.\ J.\ {\bf 172}, L95 (1972).

\bibitem{Raffaelli:2014ola} 
  B.~Raffaelli,
  Gen.\ Rel.\ Grav.\  {\bf 48},  16 (2016).

\bibitem{Abramowicz_Prasanna_1990} 
M.~A.~Abramowicz and A.~R.~Prasanna, 
Mon. Not. Roy. Astr. Soc. {\bf 245}, 720 (1990).

\bibitem{Abramowicz:1990cb} 
  M.~A.~Abramowicz,
Mon.\ Not.\ Roy.\ Astr.\ Soc.\ {\bf 245}, 733 (1990).

\bibitem{Allen:1990ci} 
  B.~Allen,
  Nature {\bf 347}, 615 (1990).

\bibitem{Hasse_Perlick_2002}
W. Hasse and V. Perlick,
Gen. Relativ. Gravit. {\bf 34}, 415 (2002).

\bibitem{Hod:2012nk} 
  S.~Hod,
  Phys.\ Rev.\ D {\bf 84}, 104024 (2011).

\bibitem{Mach:2013gia} 
  P.~Mach, E.~Malec, and J.~Karkowski,
  Phys.\ Rev.\ D {\bf 88}, 084056 (2013)

\bibitem{Chaverra:2015bya} 
  E.~Chaverra and O.~Sarbach,
  Class.\ Quant.\ Grav.\  {\bf 32}, 155006 (2015)

\bibitem{Cvetic:2016bxi} 
  M.~Cvetic, G.~W.~Gibbons, and C.~N.~Pope,
  Phys.\ Rev.\ D {\bf 94}, 106005 (2016).

 \bibitem{Koga:2016jjq} 
  Y.~Koga and T.~Harada,
  Phys.\ Rev.\ D {\bf 94}, 044053 (2016).

\bibitem{Koga:2018ybs} 
  Y.~Koga and T.~Harada,
  Phys.\ Rev.\ D {\bf 98}, 024018 (2018).

\bibitem{Koga:2019teu} 
  Y.~Koga,
  Phys.\ Rev.\ D {\bf 99}, 064034 (2019).

\bibitem{Koga:2019uqd} 
  Y.~Koga and T.~Harada,
  Phys.\ Rev.\ D {\bf 100}, 064040 (2019).

\bibitem{Keir:2014oka} 
  J.~Keir,
  Class.\ Quant.\ Grav.\  {\bf 33}, 135009 (2016).

\bibitem{Cardoso:2014sna} 
  V.~Cardoso, L.~C.~B.~Crispino, C.~F.~B.~Macedo, H.~Okawa, and P.~Pani,
  Phys.\ Rev.\ D {\bf 90}, 044069 (2014).

\bibitem{Cunha:2017qtt} 
  P.~V.~P.~Cunha, E.~Berti, and C.~A.~R.~Herdeiro,
  Phys.\ Rev.\ Lett.\  {\bf 119}, 251102 (2017).

\bibitem{Cunha:2017eoe} 
  P.~V.~P.~Cunha, C.~A.~R.~Herdeiro, and E.~Radu,
  Phys.\ Rev.\ D {\bf 96}, 024039 (2017).

\bibitem{Hod:2017zpi} 
  S.~Hod,
  Phys.\ Lett.\ B {\bf 776}, 1 (2018).

\bibitem{Visser_1995}
M. Visser,
\textit{Lorentzian Wormholes: From Einstein to Hawking} (American Institute of Physics, Woodbury, NY, 1995).

\bibitem{Morris_Thorne_1988}
M. S. Morris and K. S. Thorne,
Am. J. Phys. {\bf 56}, 395 (1988).

\bibitem{Schneider_Ehlers_Falco_1992}
P. Schneider, J. Ehlers, and E. E. Falco,
\textit{Gravitational Lenses} (Springer-Verlag, Berlin, 1992).

\bibitem{Schneider_Kochanek_Wambsganss_2006}
P. Schneider, C. S. Kochanek, and J. Wambsganss,
\textit{Gravitational Lensing: Strong, Weak and Micro,
Lecture Notes of the 33rd Saas-Fee Advanced Course},
edited by G. Meylan, P. Jetzer, and P. North (Springer-Verlag, Berlin, 2006).

\bibitem{Chetouani_Clement_1984}
L. Chetouani and G. Cl\'{e}ment,
Gen. Relativ. Gravit. {\bf 16}, 111 (1984).

\bibitem{Nandi:2006ds} 
  K.~K.~Nandi, Y.~Z.~Zhang, and A.~V.~Zakharov,
  Phys.\ Rev.\ D {\bf 74}, 024020 (2006).
  
\bibitem{Muller:2008zza} 
T.~Muller,
Phys.\ Rev.\ D {\bf 77}, 044043 (2008).

\bibitem{Tsukamoto:2016zdu}
  N.~Tsukamoto and T.~Harada,
  Phys.\ Rev.\ D {\bf 95}, 024030 (2017).

\bibitem{Tsukamoto_Harada_Yajima_2012} 
N. Tsukamoto, T. Harada, and K. Yajima,
Phys. Rev. D {\bf 86}, 104062 (2012).

\bibitem{Perlick:2014zwa} 
  V.~Perlick,
  AIP Conf.\ Proc.\  {\bf 1577}, 94 (2015).

\bibitem{Tsukamoto:2016qro} 
  N.~Tsukamoto,
  Phys.\ Rev.\ D {\bf 94}, 124001 (2016).

\bibitem{Nandi:2016uzg} 
  K.~K.~Nandi, R.~N.~Izmailov, A.~A.~Yanbekov, and A.~A.~Shayakhmetov,
  Phys.\ Rev.\ D {\bf 95}, 104011 (2017).

\bibitem{Tsukamoto:2017edq} 
  N.~Tsukamoto,
  Phys.\ Rev.\ D {\bf 95}, 084021 (2017).

\bibitem{Shaikh:2019itn} 
  R.~Shaikh, P.~Banerjee, S.~Paul, and T.~Sarkar,
  Phys.\ Rev.\ D {\bf 99}, 104040 (2019).

\bibitem{Shaikh:2019jfr} 
  R.~Shaikh, P.~Banerjee, S.~Paul, and T.~Sarkar,
  JCAP {\bf 1907}, 028 (2019).

\bibitem{Shaikh:2018oul} 
  R.~Shaikh, P.~Banerjee, S.~Paul, and T.~Sarkar,
  Phys.\ Lett.\ B {\bf 789}, 270 (2019)
  Erratum: [Phys.\ Lett.\ B {\bf 791}, 422 (2019)].

\bibitem{Muller_2004}
T.~Muller,
Am. J. Phys. {\bf 72}, 1045,(2004).

\bibitem{James:2015ima} 
  O.~James, E.~von Tunzelmann, P.~Franklin, and K.~S.~Thorne,
  Am.\ J.\ Phys.\  {\bf 83}, 486 (2015).

\bibitem{Ohgami:2015nra}
T.~Ohgami and N.~Sakai,
Phys.\ Rev.\ D {\bf 91}, 124020 (2015).

\bibitem{Ohgami:2016iqm}
T.~Ohgami and N.~Sakai,
Phys.\ Rev.\ D {\bf 94}, 064071 (2016).

\bibitem{Kuniyasu:2018cgv} 
  M.~Kuniyasu, K.~Nanri, N.~Sakai, T.~Ohgami, R.~Fukushige, and S.~Komura,
  Phys.\ Rev.\ D {\bf 97}, 104063 (2018).

\bibitem{Nambu:2019sqn} 
  Y.~Nambu, S.~Noda, and Y.~Sakai,
  Phys.\ Rev.\ D {\bf 100}, 064037 (2019).

\bibitem{Cardoso:2016rao} 
  V.~Cardoso, E.~Franzin and P.~Pani,
  Phys.\ Rev.\ Lett.\  {\bf 116}, no. 17, 171101 (2016)
  Erratum: [Phys.\ Rev.\ Lett.\ {\bf 117}, 089902 (2016)].

\bibitem{Damour:2007ap} 
  T.~Damour and S.~N.~Solodukhin,
  Phys.\ Rev.\ D {\bf 76}, 024016 (2007)

\bibitem{Lemos:2008cv} 
  J.~P.~S.~Lemos and O.~B.~Zaslavskii,
  Phys.\ Rev.\ D {\bf 78}, 024040 (2008).

\bibitem{Karimov:2019qco} 
  R.~K.~Karimov, R.~N.~Izmailov, and K.~K.~Nandi,
  Eur.\ Phys.\ J.\ C {\bf 79}, 952 (2019).

\bibitem{Bueno:2017hyj} 
  P.~Bueno, P.~A.~Cano, F.~Goelen, T.~Hertog, and B.~Vercnocke,
  Phys.\ Rev.\ D {\bf 97}, 024040 (2018).

\bibitem{Volkel:2018hwb} 
  S.~H.~Volkel and K.~D.~Kokkotas,
  Class.\ Quant.\ Grav.\  {\bf 35}, 105018 (2018).

\bibitem{Paul:2019trt}
S.~Paul, R.~Shaikh, P.~Banerjee, and T.~Sarkar,
JCAP {\bf 03}, 055 (2020).

\bibitem{Amir:2018pcu} 
  M.~Amir, K.~Jusufi, A.~Banerjee, and S.~Hansraj,
  Class.\ Quant.\ Grav.\  {\bf 36}, no. 21, 215007 (2019).

\bibitem{Nandi:2018mzm} 
  K.~K.~Nandi, R.~N.~Izmailov, E.~R.~Zhdanov, and A.~Bhattacharya,
  JCAP {\bf 1807}, 027 (2018).

\bibitem{Ovgun:2018fnk} 
  A.~Ovgun,
  Phys.\ Rev.\ D {\bf 98}, 044033 (2018).

\bibitem{Bhattacharya:2018leh} 
  A.~Bhattacharya and R.~K.~Karimov,
  arXiv:1811.00768 [gr-qc].
  
\bibitem{Ovgun:2018swe} 
  A.~Ovgun,
  arXiv:1811.06870 [gr-qc].

\bibitem{Ovgun:2018oxk} 
  A.~Ovgun,
  Universe {\bf 5}, 115 (2019).
  
\bibitem{Tsukamoto:2019ihj} 
  N.~Tsukamoto and T.~Kokubu,
  Phys.\ Rev.\ D {\bf 101}, 044030 (2020).

\bibitem{Tsukamoto:2016jzh} 
  N.~Tsukamoto,
  Phys.\ Rev.\ D {\bf 95}, 064035 (2017).

\bibitem{Eiroa:2002mk} 
  E.~F.~Eiroa, G.~E.~Romero, and D.~F.~Torres,
  Phys.\ Rev.\ D {\bf 66}, 024010 (2002).

\bibitem{Bozza:2005tg} 
  V.~Bozza, F.~De Luca, G.~Scarpetta, and M.~Sereno,
  Phys.\ Rev.\ D {\bf 72}, 083003 (2005).
  [gr-qc/0507137].

\bibitem{Bozza:2006nm} 
  V.~Bozza, F.~De Luca, and G.~Scarpetta,
  Phys.\ Rev.\ D {\bf 74}, 063001 (2006).

\bibitem{Bozza:2007gt} 
  V.~Bozza and G.~Scarpetta,
  Phys.\ Rev.\ D {\bf 76}, 083008 (2007).

\bibitem{Ishihara:2016sfv} 
  A.~Ishihara, Y.~Suzuki, T.~Ono, and H.~Asada,
  Phys.\ Rev.\ D {\bf 95}, 044017 (2017).

\bibitem{Wei:2011zw} 
  S.~W.~Wei, Y.~X.~Liu, and H.~Guo,
  Phys.\ Rev.\ D {\bf 84}, 041501 (2011).

\bibitem{Stefanov:2010xz} 
  I.~Z.~Stefanov, S.~S.~Yazadjiev, and G.~G.~Gyulchev,
  Phys.\ Rev.\ Lett.\  {\bf 104}, 251103 (2010).

\bibitem{Iyer:2006cn}
S.~V.~Iyer and A.~O.~Petters,
Gen. Rel. Grav. {\bf 39}, 1563 (2007).

\bibitem{Bozza:2008ev} 
  V.~Bozza,
  Phys.\ Rev.\ D {\bf 78}, 103005 (2008).

\bibitem{Tsukamoto:2014dta}
N.~Tsukamoto, T.~Kitamura, K.~Nakajima, and H.~Asada,
Phys. Rev. D \textbf{90}, 064043 (2014).

\bibitem{Weinberg:1972kfs} 
  S.~Weinberg,
  ``Gravitation and Cosmology : Principles and Applications of the General Theory of Relativity,''
  John Wiley \& Sons, New York, USA, 1972.

\bibitem{Poisson}
E. Poisson, 
\textit{A Relativist's Toolkit: The Mathematics of Black-Hole Mechanics,}
(Cambridge University Press, Cambridge, 2004). 


\end{thebibliography}
\end{document}